\documentclass[aps,prl,twocolumn,superscriptaddress,nofootinbib,longbibliography]{revtex4-2}

\usepackage{amsmath,amssymb}
\usepackage{amsfonts}
\usepackage[dvipsnames]{xcolor}
\usepackage{graphicx}
\usepackage{dcolumn}

\usepackage{epstopdf}
\usepackage{color}
\usepackage[utf8]{inputenc}
\usepackage{amsthm}
\usepackage{fontenc}
\usepackage{soul}

\usepackage{bm}
\RequirePackage{slashed}\usepackage{cancel}
\usepackage[normalem]{ulem}

\def\sumint{\int \! \!\ \! \! \! \! \!\ \! \! \!\! \!\sum}
\def\be{\begin{eqnarray} &&} 
\def\nonu{\nonumber \\ &&} 
\def\ee{\end{eqnarray}} 
\usepackage{array,booktabs}
\newcolumntype{C}{>{$\displaystyle}c<{$}}

\begin{document}

\pacs{13.60.Hb,14.20.Dh,27.10.+h}

\title{The European Muon Collaboration effect in Light-Front Hamiltonian Dynamics}

\author{Emanuele Pace}
\affiliation{Universit\`a di Roma ``Tor Vergata'', 
Via della Ricerca Scientifica 1, 00133 Rome, Italy}
\author{Matteo Rinaldi} 
\affiliation{ Dipartimento di Fisica e Geologia,
Universit\`a degli Studi di Perugia and Istituto Nazionale di Fisica Nucleare,
Sezione di Perugia, via A. Pascoli, I - 06123 Perugia, Italy}
\author{Giovanni Salm\`e}
\affiliation{Istituto  Nazionale di Fisica Nucleare, Sezione di Roma, Piazzale A. Moro 2,
00185 Rome, Italy} 
\author{Sergio Scopetta}
\affiliation{ Dipartimento di Fisica e Geologia,
Universit\`a degli Studi di Perugia and Istituto Nazionale di Fisica Nucleare,
Sezione di Perugia, via A. Pascoli, I - 06123 Perugia, Italy}


\begin{abstract}
A rigorous light-front formalism for electron deep inelastic scattering on unpolarized nuclei, in Bjorken limit, is reported. It preserves Poincar\'e covariance, macroscopic locality, both number of particles and momentum sum rules. The scheme is applied to the A=3 iso-doublet, very relevant 
 {in view of the planned operation with   unpolarized and polarized beams at the Electron-Ion Collider}. At variance with previous light-front estimates, our procedure, including a realistic nuclear description and free-nucleon structure functions, predicts a sizeable European Muon Collaboration effect for $^3$He. This will allow to analyze deviations from the proposed baseline in terms of genuine QCD effects. The extension to heavier nuclei is straightforward, although numerically challenging.
\end{abstract}

\maketitle

After almost forty years of studies, the European Muon Collaboration (EMC) effect \cite{EuropeanMuon:1983wih}, with the characteristic depletion of the ratio  $R^A_{EMC}(x)$ between electron-nucleus and electron-deuterium deep inelastic scattering (DIS) cross sections (normalized to the relative number of nucleons) {in the intermediate region of the Bjorken variable $x$},
has not yet been fully understood.
Any attempt to explain the electromagnetic (em) response of the nucleus in the DIS kinematic region simply as the incoherent response of the {\em constituent} nucleons, assumed to be the only degrees of freedom (dof) acting in the nucleus, failed to succeed.
 {This 
finding  raised the question about how and to what extent the nucleus can be considered as a QCD laboratory, where  quark and gluon dof's  play a significant role.
Unfortunately, a quantitative answer is still lacking, although important progresses have been made, 
 e.g. by determining the role of the short-range correlations between the nucleons once the nucleus is probed by  high-virtuality photons  \cite{Hen:2016kwk,CLAS:2019vsb,Wang:2020uhj}, and also  by addressing the medium modification of the nucleons, given the large overlap between them (see, e.g. Ref. \cite{Cloet:2019mql})}.
Perhaps, the answer could be found in the study of complementary semi-inclusive
and exclusive
processes \cite{Dupre:2015jha},
 planned for light nuclei
at existing 
\cite{Armstrong:2017zqr}
and future \cite{AbdulKhalek:2021gbh} high-energy facilities, like
 the  Electron-Ion Collider (EIC), where a relevant program
of measurements with (polarized) $^3$He beams is under development. For now, new $^3$He and $^3$H DIS data have been taken  at JLab by the MARATHON Collaboration \cite{MARATHON:2021vqu} 
{allowing a fresh extraction of} the ratio of the neutron to proton structure functions 
{(SFs)}
{$r(x) = F_2^n(x) / F_2^p(x)$, 
needed for the flavor decomposition 
{(see also Ref. \cite{Cocuzza:2021rfn})}. 
{This experimental scenario} has motivated our efforts to improve the quantitative analysis of the EMC effect for the bound three-nucleon systems.

 In order to make highly reliable
 search and {discovery} of  effects that can  explain the EMC results, one should first establish a {sound} baseline, which means i) embedding general principles to the greatest  extent and ii) including  the very successful description of nuclei, elaborated by phenomenologists  over decades. In view of this,
we propose
a Poincar\'e-covariant formalism for nuclear DIS
 without medium deformations
of the bound nucleons, but 
{fully able to exploit}
the sophisticated description of nuclei, as achieved nowadays. The approach, to be applied to the nuclear SFs, has been carefully developed for $^3$He  in Ref. \cite{Pace:2020ned} within the light-front Hamiltonian dynamics (LFHD) \cite{Dirac:1949cp,KP} (the most suitable when an em probe is used), and it is valid for any nucleus. { A Poincar\'e-covariant description of nuclei and in particular of $^3$He will have a remarkable impact on the interpretation of the experimental results at the EIC, where a crucial part of the program will involve light-ion beams (see, e.g., sections 2.6, 2.7, 7.2, 7.3 of Ref. \cite{AbdulKhalek:2021gbh}).}

The main ingredient is the light-front (LF) spectral function
\cite{DelDotto:2016vkh}, ${\cal P}^{N}(\tilde{\bm \kappa},\epsilon)$,
the probability
distribution of finding
a nucleon with
LF momentum
$\tilde{\bm \kappa}$ 
\footnote{
The light-front components of a four vector $v$ are $(v^-,{\bf \tilde  v})$, where
${\bf \tilde  v}=(v^+,{\bf v}_ \perp)$ with $v^\pm=v^0 \pm {\bf {\hat {n}}} \cdot {\bf v}$ and 
${\bf v}_ \perp= {\bf v} - {\bf {\hat {n}}} ({\bf {\hat {n}}} \cdot {\bf v})$. The vector 
${\bf {\hat {n}}}$ is a generic unit vector. In this paper we choose ${\bf {\hat {n}}} \equiv {{\hat {z}}}$.}
in the intrinsic reference frame of the cluster $[1, (A-1)]$ and the fully interacting $(A-1)$-nucleon system
with
 intrinsic energy
 $\epsilon$.
{In particular, notice that we use} nonsymmetric intrinsic variables,
{which disentangle the internal motion of the interacting $(A-1)$-nucleon system from 
{that} of the 
{struck} nucleon and hence are able to implement the macroscopic locality  \cite{KP}.} 
{Recently,} in this framework, the six {$^3$He} leading-twist transverse-momentum distributions 
have been evaluated
in  valence approximation \cite{Alessandro:2021cbg}.

In what follows,   
{adopting the Bjorken limit, we first introduce the} unpolarized
nuclear SF, $F^A_2(x)$, 
(see also  Ref. \cite{Pace:2020ned}), and show
novel spin-independent and spin-dependent light-cone momentum distributions.
To evaluate the nuclear SF, 
{according to our final goal, we have to adopt the free-nucleon SFs}, for which many parametrizations exist. To minimize the dependence upon the chosen
nucleon SFs { and test the self-consistency of our approach,} we  extract    $r(x)$ from  the experimental 
   DIS cross-sections off $^3$He and $^3$H nuclei
    \cite{MARATHON:2021vqu}, {by applying
the iterative method of Refs. \cite{Pace:2000ky,Pace:2001cm} and  our Poincar\'e-covariant expression of the SF $F^{A=3}_2$}.  Finally from 
    $r(x)$ and  {the proton SF} $F_2^p(x)$ of Ref. \cite{SpinMuonSMC:1997voo}
  the EMC ratio, $R^{A=3}_{EMC}(x)$,  is evaluated and compared with existing experimental data.
  
It is very important to  notice that previous calculations elaborated within the LF dynamics did not predict any EMC effect, i.e. it was obtained $R^A_{EMC}(x)\ge 1$,
 as
discussed, e.g., in Refs. 
\cite{Smith:2002ci,Miller:2001tg}. More precisely, it is generally thought that the EMC effect  cannot be explained by  unmodified nucleons {only}
(see, e.g., Refs. \cite{Hen:2016kwk,CLAS:2019vsb}, for the in-medium modifications). However, 
{notice} that  in Refs. \cite{Smith:2002ci,Miller:2001tg} 
a nuclear mean field is used, so that the nuclear correlations are not properly treated.
 In addition,  in Refs. \cite{Oelfke:1990uy,Coester:1992cg} 
  LF calculations {with symmetric intrinsic variables}
{were} {{performed and covariance was only assured for kinematical Lorentz transformations}}.
Therefore, accurate studies of the EMC  effect within a Poincar\'e-covariant framework, enriched by a sound dynamical content, are mandatory.
{For the latter purpose}, in this work 
{the three-body wave functions (wfs)}
\cite{Kievsky:1994mxj,Kievsky:1995uk}, obtained from 
the NN charge-dependent Av18  potential \cite{Wiringa:1994wb} (including the Coulomb interaction) plus  the NNN Urbana IX force \cite{Pudliner:1995wk}, are used.

{\it Formalism.}
 We adopt the  reference frame   
{{where the nucleus three-momentum}} {is}
${\bf P}_{A} = 0$ and the four-momentum transfer has components:  $ q\equiv\{q_0,
q_z=-  {|{\bf q}|},{\bf 0}_\perp\}$.
{Moreover, we assume the impulse approximation (IA), that is  suitable for describing a high-virtuality photon that impinges on a nucleon inside the nucleus,  leaving unaffected the fully interacting spectator system. 
In Ref.  \cite{Lev:1998qz}  it was shown that the 
{IA} satisfies the covariance with respect to Poincar\'e transformations
plus time reversal and parity, if applied in the Breit frame with the momentum transfer ${\bf q}_\perp=0$ and in any frame 
{reachable} through a LF boost parallel to the $z$ axis. 
{In} IA, 
 {one} obtains the following general expression of the nuclear SF   
   (see Ref. \cite{Pace:2020ned})}
\be
\hspace{-4mm} F^A_2(x) =  
   \sum_N 
\hspace{-0.5mm}
\sumint {d\epsilon }
\hspace{-1mm}
  \int \hspace{-1mm}{d{\bf 
 k }_{\perp} \over (2 \pi)^3}
 \int_{\xi^B_{min}}^{1} \hspace{-3mm} {d\xi } 
~
  {{\cal P}^{N}(\tilde{\bm \kappa},\epsilon) 
  \hspace{0.5mm} 
  E_S \over 2~ \kappa ^+ (1- \xi)}  
  F^N_2(z)  ,
   \label{F2b}
\ee
where 
  i) $\tilde{\bm \kappa}\equiv\{\kappa ^+,{\bf k }_{\perp}\}$, 
  with
$\kappa ^+ = \xi {\cal M}_0(1,A-1)$ and  ${\cal M}_0(1,A-1)$ the free mass of the cluster 
$[1,(A-1)]$; ii) $E_S=\sqrt{{M^2_S}+|{\bf k}_\perp|^2+|{\bm \kappa}_z|^2}$ is  the total energy of the  $(A-1)$-nucleon spectator  system  (see Eq. (46) in Ref. \cite{DelDotto:2016vkh} for the expression of $\kappa_z$), with  $M_{S}= (A-1)m +\epsilon$  the mass of the interacting $(A-1)$-nucleon system,   $\epsilon$ {its} intrinsic energy and $m$ the nucleon mass; {iii) $z = Q^2 / (2 p \cdot q)$ with $p$ the off-shell nucleon four-momentum.

{In the Bjorken limit, one} can readily exchange the integrations on $\epsilon$ and ${\bf k}_{\perp}$ and the integration on $\xi$,
since both 
$z= {\xi^B_{min}/ \xi }$ and $\xi^B_{min} = ~ x ~{m / M_A}$ do not depend on $\epsilon$ and ${\bf k}_{\perp}$.  
Then the nuclear SF becomes 
\be
 F^A_2(x) =  ~  \sum_N ~
    \int_{\xi^B_{min}}^{1}  d\xi  ~ 
  F^N_2(z)
  ~ {{f}}^N_1(\xi) ~ ,
 \label{F2a}
\ee
where ${{f}}^N_1(\xi)$ is the unpolarized light-cone momentum distribution given by
\be
{{f}}^N_1(\xi) =  \int d{\bf k}_{\perp} ~ n^N(\xi, {\bf k}_{\perp})~,
 \label{F2ab}
 \ee
with
    $n^N(\xi, {\bf k}_{\perp})$ 
   the LF spin-independent nucleon momentum distribution
   (see Refs. \cite{DelDotto:2016vkh,Alessandro:2021cbg}), 
    viz.
\be
 n^N(\xi, {\bf k}_{\perp}) = \sumint {d\epsilon } ~
  {{\cal P}^{N}(\tilde{\bm \kappa},\epsilon) \over (2 \pi)^3~ 2~ \kappa ^+} 
  ~
  {E_S \over  (1- \xi)} \, .
   \label{F2aa}
   \ee
   {This quantity} is the trace of 
the momentum-distribution $2\times 2$ matrix, ${\cal N}^N_{\cal M}(\xi,{\bf k}_\perp, {\bf S})$ {\cite{Alessandro:2021cbg}}, with ${\bf S}$   the polarization of the nucleus and ${\cal M}$ the  component of the total angular momentum ${\cal J}$ 
along $\bf S$
 (in general ${\bf S}\ne \hat z$). The trace is obviously independent of $\bf S$ and ${\cal M}$.
It is worth noticing that, in our Poincar\'e-covariant framework, the momentum distribution 
  fulfills  both normalization  (i.e. the baryon number sum rule) and momentum sum rule
 \cite{DelDotto:2016vkh,Alessandro:2021cbg}. Finally, in the Bjorken limit, the evaluation of the nuclear SF is simplified, since  $n^N(\xi, {\bf k}_{\perp})$ 
can be obtained  more directly from  the nucleus wf in momentum space \cite{Alessandro:2021cbg},
rather  than through the cumbersome  LF spectral-function. Hence, once the ground-state of a nucleus is  calculated {with a realistic interaction}, the corresponding nuclear structure {function}  can be affordably 
 {evaluated} in the Bjorken limit, through Eqs. \eqref{F2a} and \eqref{F2ab}. Differently, for describing the  nuclear DIS  in 
IA, one needs the knowledge of the  LF spectral function ${\cal P}^{N}(\tilde{\bm \kappa},\epsilon)$
\cite{Pace:2020ned}.

For $A=3$, Eq. (4), in terms of the
wf in momentum space, ~ ${\bf {\cal G}}_{{L_{\rho}}X}^{j_{23} l_{23}s_{23}}(k_{23},k)$,  reads
\cite{Alessandro:2021cbg}
\be
 n^N(\xi, {\bf k}_{\perp}) 
 =  \sum_{T_{23},\tau_{23}}  
 \langle T_{23} \tau_{23} {1 \over 2} \tau |{1 \over 2} T_z \rangle^2
  \int_0^\infty  d { k}_{23}~ k^2_{23}
\nonu\times    
{E_{23}~E({\bf k}) \over 4 \pi ~(1- \xi) k^+} \hspace{-0mm}
\sum_{j_{23},l_{23},s_{23}}  \sum_{L_{\rho},X} ~ 
  |{\bf {
  \cal 
  G}}_{{L_{\rho}}X}^{j_{23} l_{23}s_{23}}(k_{23},k)|^2 
  ,
 \label{F2e}
\ee
where ${\bf k}_{23}$ is the Cartesian momentum for the internal motion of the [23]-pair
 and ${\bf k}$ the one of the probed nucleon in the {intrinsic} frame of the $A=3$ system, 
$E_{23}=\sqrt{M^2_{23}+{{\bf k}}^2 }$,
with  $M_{23}=2~\sqrt{m^2+|{\bf k}_{23}|^2}$, $k^+ = \xi M_0(1,2,3)  \ne \kappa^+$, with $M_0(1,2,3)$ the free mass of the $A=3$ system, and  $E({\bf k})= \sqrt{m^2+|{\bf k}|^2 }$. The  nonsymmetric intrinsic variables ${\bf k}$ and ${\bf k}_{23}$ (see \cite{DelDotto:2016vkh} for their definition) generate the relevant 
coefficients appearing in the integrand.

For  $A=3$  one can immediately calculate
 the spin-independent LF-momentum distribution in Eq. (\ref{F2e}), and then the light-cone momentum
distributions, ${{f}}^N_1(\xi)$, Eq. \eqref{F2ab}.
Analogously, one can evaluate the  longitudinally and transversely polarized nucleon distributions,
$g_1^N$ and $h_1^N$, by  using the spin-dependent nucleon momentum distribution, $n^N_{\sigma,\sigma'}(\xi,{\bf k}_\perp, {\bf S})$, i.e. the  elements of ${\cal N}_{\cal M}^N(\xi, {\bf k}_\perp,{\bf S})$ \cite{Alessandro:2021cbg}. 
These distributions are given by 
\be
{{g}}^N_1(\xi) =  \int d{\bf k}_{\perp} ~ \Delta f^N(\xi, {\bf k}_{\perp})
\, ,
 \label{g1ac}
 \ee
 \be
{{h}}^N_1(\xi) =  \int d{\bf k}_{\perp} ~ \Delta_T' f^N(\xi, {\bf k}_{\perp})
 \, ,
 \label{h1ac}
 \ee 
 where $\Delta f^N(\xi, {\bf k}_{\perp})$ and 
$\Delta_T' f^N(\xi, {\bf k}_{\perp})$ are two out of the six leading-twist transverse-momentum distributions {\cite{Alessandro:2021cbg}}. 
In terms of  ${\cal N}^N_{\cal M}(\xi, {\bf k}_\perp, {\bf S} )$ {elements}  (for ${\cal J}=1/2$ the dependence upon ${\cal M}$ amounts to a trivial phase~\cite{Alessandro:2021cbg}),  one gets 
 \be
g^N_1(\xi) =  
\int d{\bf k}_{\perp}  \Bigl[n^N_{++}( \xi, {\bf k}_{\perp},\hat z)-n^N_{--}( \xi, {\bf k}_{\perp},\hat z)\Bigr]
 ~,
 \label{g1ab}
 \ee
 and
 \be
{{h}}^N_1(\xi) 
=\int d{\bf k}_{\perp} \Bigl[n^N_{+-}(\xi, {\bf k}_{\perp},{\hat x})+n^N_{-+}(\xi, {\bf k}_{\perp},{\hat x})\Bigr]
~ ,
 \label{h1ab}
 \ee
where {longitudinal and transverse-spin states are used, and one should recall that ${\bf S}~||~\hat {\bf z}$ for $g_1^N$ and  ${\bf S}\perp\hat {\bf z}$ for $h_1^N$.}

 The distributions $g^N_1(\xi)$ and $h^N_1(\xi)$ are crucial for{: i)}  evaluating inclusive spin-dependent SFs {needed} 
 in the IA treatment of polarized DIS  and devising a proper extraction method for collecting the neutron information from the $^3$He nucleus
  (see, e.g., Ref. \cite{CiofidegliAtti:1993zs} for the {nonrelativistic (NR)} case), and ii) {analysing} semi-inclusive DIS 
 off transversely polarized $^3$He (see Refs. \cite{Scopetta:2006ww,Kaptari:2013dma} for the NR case), {so that}
 the neutron single-spin asymmetries {can be obtained}. 
 Notice that {the} relativistic evaluation {of these distributions} is particularly relevant in view of the future measurements at high energy facilities, such as the EIC \cite{AbdulKhalek:2021gbh}.
 
{\it Results.} For the first time, 
the nucleon momentum distributions given in Eqs. \eqref{F2ab}, \eqref{g1ab} and \eqref{h1ab} are evaluated  for $^3$He  within a fully Poincar\'e-covariant approach, {by using  wfs obtained 
{along the line of} Refs.
\cite{Kievsky:1994mxj,Kievsky:1995uk}}, and are shown in Figs. \ref{fig_f1}, \ref{fig_g1}  and \ref{fig_h1}, respectively. 
The difference between the {longitudinally and transversely polarized} distributions is a measure of the relevance of the relativistic effects, which appear to be 
more sizable for the proton than for the neutron, 
{given that only small partial waves with high orbital angular momentum
contribute to the tiny proton polarization in $^3$He}. {As expected, $g_1^n$ and $h_1^n$ are remarkably larger than the corresponding proton distributions.}

\begin{figure}[htb]
\begin{center}
\includegraphics[width=9. cm]{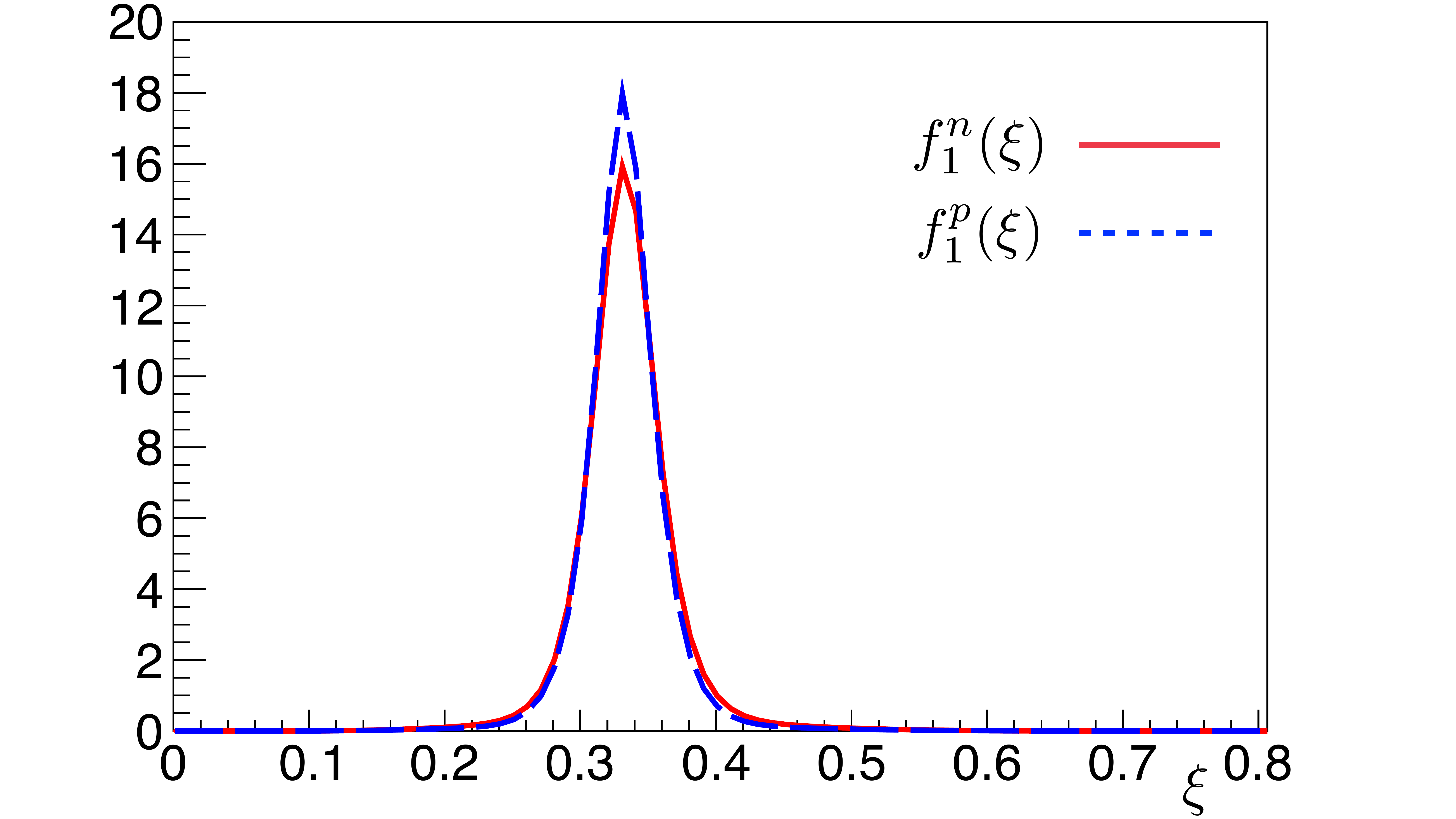}
\caption{(Color online). Light-cone momentum
distributions, {{Eq. (\ref{F2ab})}} for proton and neutron in $^3$He.} 
\label{fig_f1}
\end{center}
\end{figure}
\begin{figure}[htb]
\begin{center}
\includegraphics[width=9. cm]{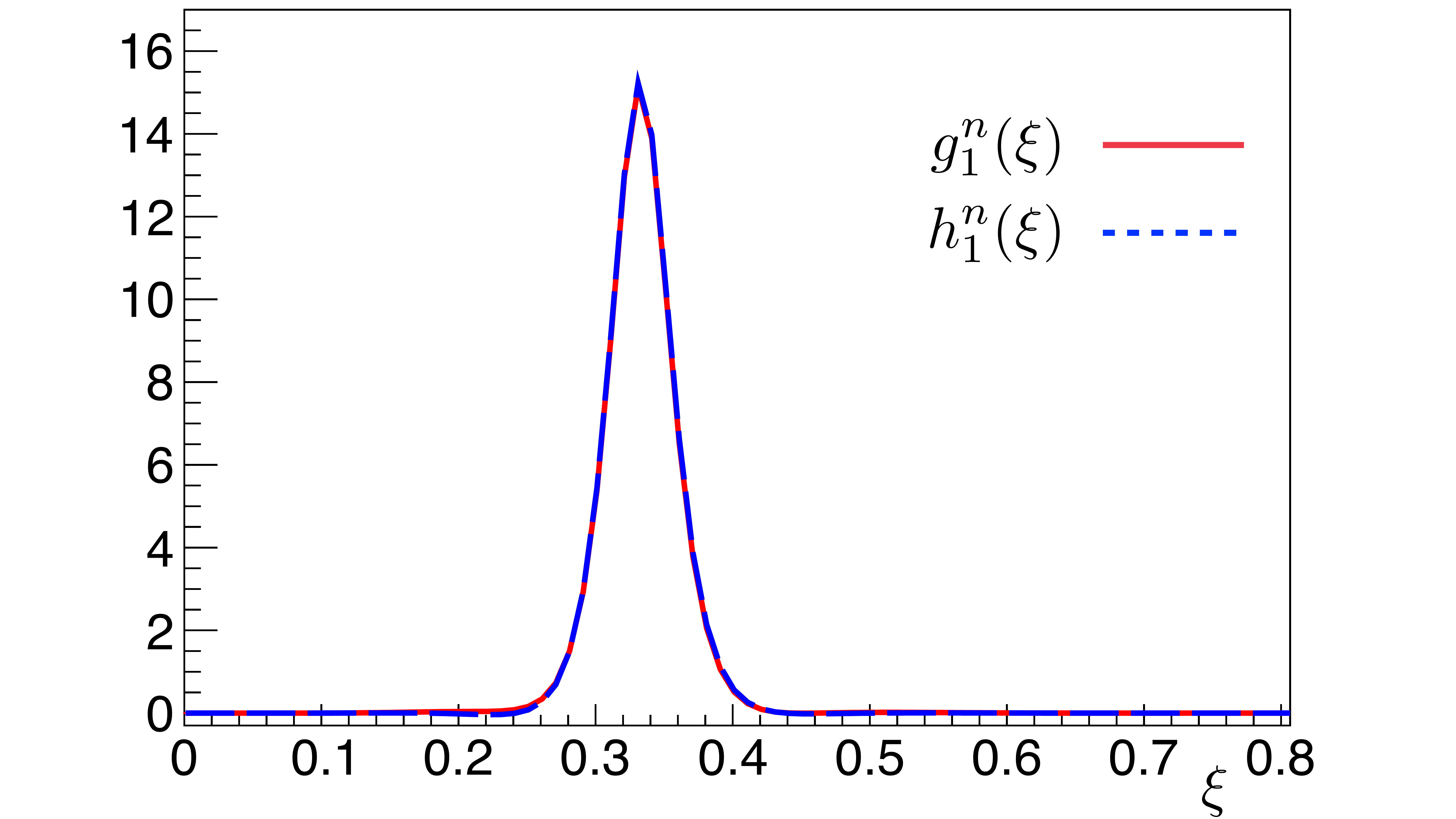}
\caption{(Color online). Light-cone helicity and
transversity momentum
distributions, {{Eqs. (\ref{g1ab}) 
and (\ref{h1ab})}}, for a neutron in $^3$He.} 
\label{fig_g1}
\end{center}
\end{figure}
\begin{figure}[htb]
\begin{center}
\includegraphics[width=9. cm]{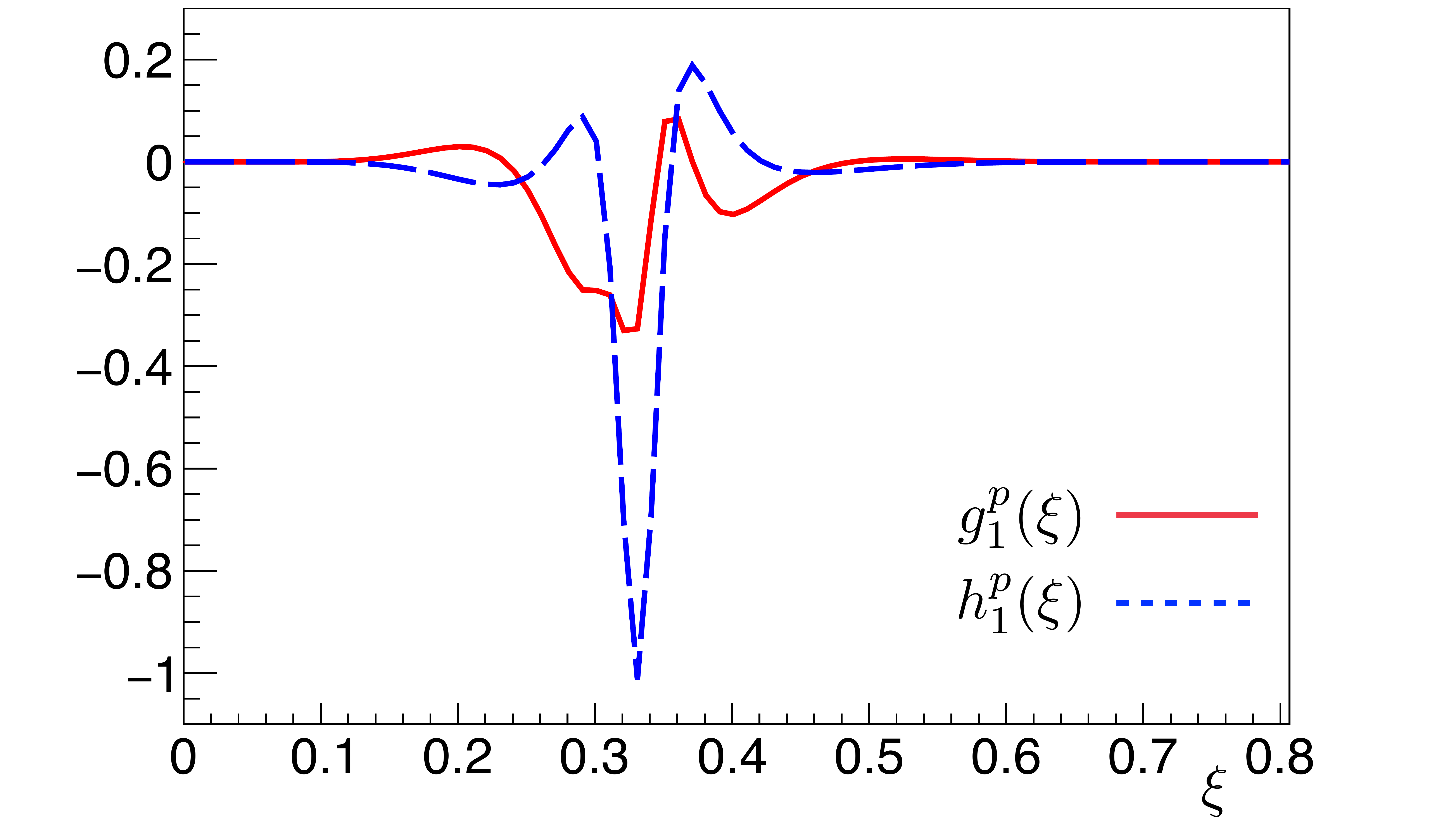}
\caption{(Color online). As in Fig. \ref{fig_g1}, but for a proton in $^3$He.} 
\label{fig_h1}
\end{center}
\end{figure}
To evaluate the nuclear SF from Eq. (\ref{F2a}) both 
$F_2^p$ and $F_2^n$ are needed. Many parametrizations 
exist
for $F_2^p$ (see, e.g., {Refs. \cite{SpinMuonSMC:1997voo, EuropeanMuon:1987obv,Gluck:1991ng,Martin:2001es}}) from proton experimental data, while $F_2^n$ is elusive, due to the lack of
{free} neutron targets. {Thus, for $F^n_2(x)$ we largely rely  on experimental information provided by the new accurate data, $E^{ht}(x)=\sigma^h / \sigma^t$, from  Ref. \cite{MARATHON:2021vqu}.}
{We also take into account  the two  results at $x=0$ and $x=1$ (with large uncertainties) obtained in Ref. \cite{Cui:2021gzg} applying a careful extrapolation method to $E^{ht}(x)$ \cite{MARATHON:2021vqu}.
{Then we perform a quadratic least-squares interpolation of  the experimental quantity $E^{ht}(x)$ \cite{MARATHON:2021vqu} plus the two extrapolated values \cite{Cui:2021gzg}, obtaining the function ${\cal E}^{ht}(x)$ }  (cf. Fig. 1 in the Supplemental Material), which is necessary  for}
adopting the iterative method presented in Refs. \cite{Pace:2000ky,Pace:2001cm} and  eventually   extracting $r(x)=F_2^n(x)/F_2^p(x)$. 

Let us define the 
ratio
 \be
R^A_2(x)={F^A_2(x)\over Z~F^p_2(x)+(A-Z)~F^n_2(x)} ~,
\label{isosp}
\ee
and the super-ratio 
 \be
R^{ht}(x)={ R^{^3He}_2(x) \over R^{^3H}_2(x)} = E^{ht}(x) {2 r(x) + 1 \over r(x)+2 } ~ .
\label{super}
\ee
Noteworthy, in IA the super-ratio is a functional of $r(x)$ 
($R^{ht}(x) = R^{ht}[x,r(x)]$), once 
a fixed parametrization for $F^p_2(x)$ is used. 
Hence, the iterative method \cite{Pace:2000ky,Pace:2001cm} can be applied for determining  the ratio $r(x)$, i.e.
 \be
r^{(n+1)}(x) = -~{{\cal E}^{ht}(x) - 2 R^{ht}[x,r^{(n)}(x)] \over
2 {\cal E}^{ht}(x)-R^{ht}[x,r^{(n)}(x)]  }~,
\label{recur}
\ee
where $R^{ht}[x,r^{(n)}(x)]={ R^{^3He}_2[x,r^{(n)}(x)] / R^{^3H}_2[x,r^{(n)}(x)]}$ is obtained from our
Poincar\'e-covariant calculations of Eq. \eqref{F2a} for the $A=3$ nuclei.

{After assigning  the  initial $r^{(0)} ( x )$, in the present calculation this quantity  is built from  $F_2^{n(p)}$ of Ref. \cite{SpinMuonSMC:1997voo}, one gets   the results  shown in Fig. \ref{fig:rx} (notice that  by using  $F_2^p(x)$ of Refs.  \cite{EuropeanMuon:1987obv,Gluck:1991ng,Martin:2001es} a similar final  $r(x)$ is achieved). There, the full dots represent  stable outcomes, i.e. with a relative difference between the calculated ratios after 100 and 200 iterations less than $4$\textperthousand. The uncertainties associated to our result (the shaded area) are attained by applying to the full dots both a quadratic least-squares interpolation and a cubic one, that corresponds to a better fit (see in the Supplemental Material the actual expressions).
Moreover, in Fig.  \ref{fig:rx},   our $r(x)$ is compared
with i) the data extracted by MARATHON \cite{MARATHON:2021vqu} through  Eq. \eqref{super} with  the theoretical quantity $R^{ht}(x)$ evaluated following Refs. \cite{Kulagin:2004ie,Kulagin:2010gd}, and ii) the SMC $r(x)$ ratio \cite{SpinMuonSMC:1997voo}.}
\begin{figure}[htb]
\begin{center}
\includegraphics[width=9.2 cm]{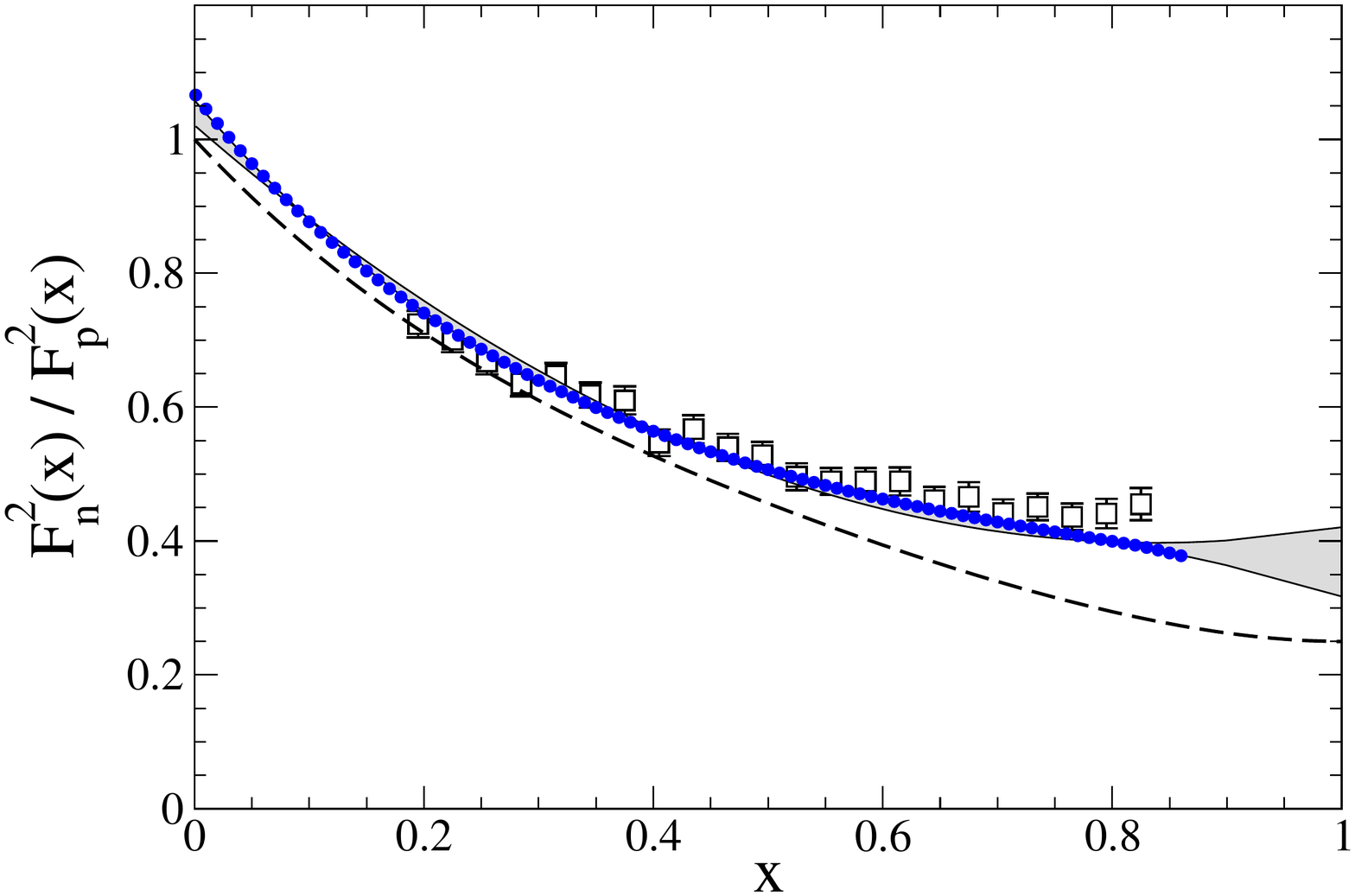}
\caption{The ratio $r(x)=F_2^n(x)/F_2^p(x)$ obtained through the recurrence relation of Eq. \eqref{recur},  full dots (see text for the $x$-range),  compared with i) the 
{results} extracted in Ref. \cite{MARATHON:2021vqu}, empty squares, and  ii) the the ratio from the SMC $F^{p(n)}_2(x)$  \cite{SpinMuonSMC:1997voo}, dashed line. The shaded area is the uncertainty obtained by using i) a quadratic least-squares interpolation of the full-dot set and ii) a cubic one (see text).
}
\label{fig:rx}
\end{center}
\end{figure}

Once $r(x)$ is determined, the nuclear structure  functions of the iso-doublet,  Eq. \eqref{F2a},  are evaluated by using i)  the  parametrization for  $F_2^p(x)$  of Ref. \cite{SpinMuonSMC:1997voo},
ii) $F_2^n(x) = r(x) F_2^p(x)$, and iii) the light-cone distributions from Eq. \eqref{F2ab}. In particular, after specializing to $A=3$ and $A=2$ the ratio in Eq. \eqref{isosp},
 one builds the EMC ratio
for the three-body system, viz. 
   \be
R^{^3{\rm He}(^3\rm H)}_{EMC}(x)={R^{^3{\rm He}(^3{\rm H})}_2(x)\over R^{\rm D}_2(x)}  ~ .
\label{rat}
\ee
 In Fig. \ref{fig:EMC}  the comparison  between the EMC experimental data for  $^3$He \cite{Seely:2009gt}, rescaled according to Ref. \cite{Kulagin:2010gd} (see also Ref. \cite{MARATHON:2021vqu})
  and the outcomes of our Poincar\'e-covariant approach (cf. Eq. \eqref{F2a}) is presented. The deuteron ratio $R^{\rm D}_2(x)$ has been evaluated by using the  wf corresponding to the  NN AV18 interaction and 
 {the expression of the LF deuteron SF 
 consistent with Eq. \eqref{F2a} for $A=2$ (see Ref. \cite{Pace:2001vg}). For the $^3$He ratio, in order to study the dependence upon the nuclear interaction, we have adopted i) the $^3$He wf of Refs.~\cite{Kievsky:1994mxj,Kievsky:1995uk}, corresponding to the charge dependent NN AV18~\cite{Wiringa:1994wb} interaction plus the NNN Urbana IX force~\cite{Pudliner:1995wk}, and ii) the one calculated without the NNN potential.}
  
\begin{figure}[htb]
\begin{center}
\includegraphics[width=9.cm]{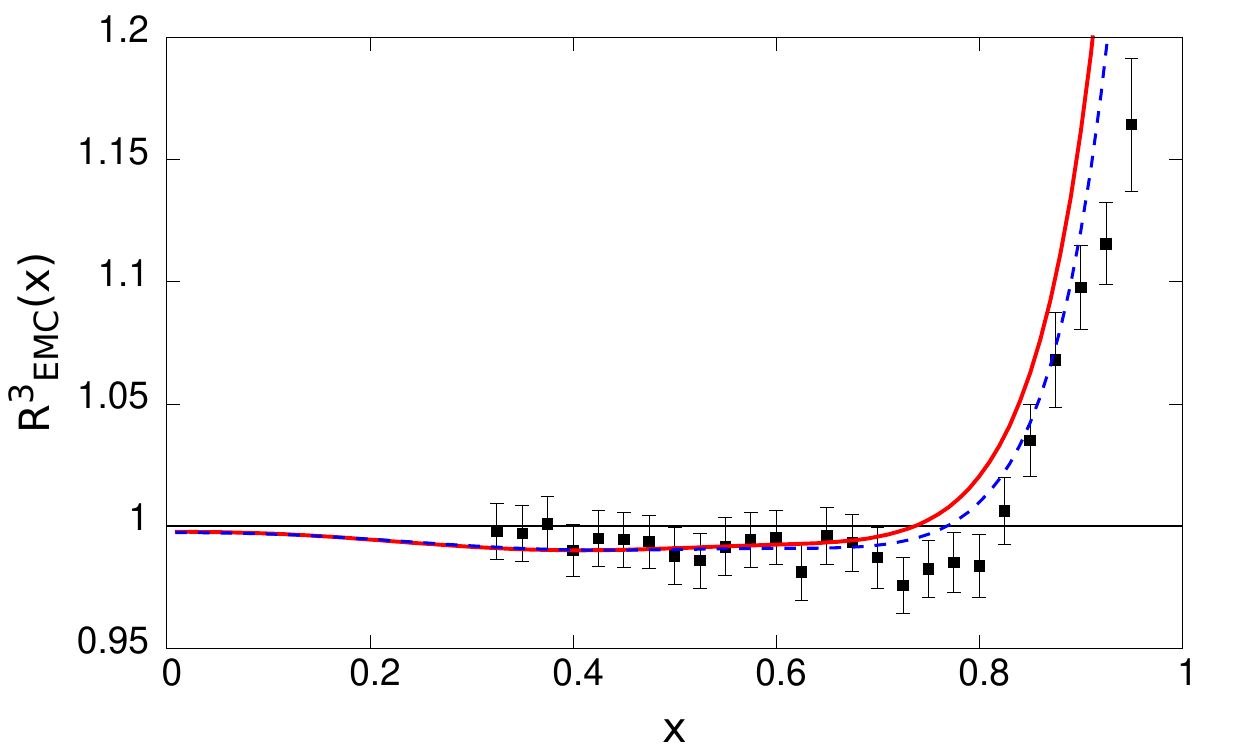}
\caption{
The EMC ratio in $^3$He. Solid line: our calculation with i) $F_2^p(x)$  of Ref.  \cite{SpinMuonSMC:1997voo},  ii) $F_2^n(x) = r(x) F_2^p(x)$,  where  $r(x)$ has been determined through Eq. \eqref{recur} and iii) the  light-cone distributions  corresponding to the NN AV18~\cite{Wiringa:1994wb} interaction plus the NNN Urbana IX force~\cite{Pudliner:1995wk}.  
Dashed line: the same as the solid line, but excluding the NNN potential. Black squares: Jlab  data of Ref. \cite{Seely:2009gt}, 
as reanalysed 
in Ref. \cite{Kulagin:2010gd}. The uncertainties on $r(x)$, shown in Fig. \ref{fig:rx}, are fully absorbed  by the width of the lines.}
\label{fig:EMC}
\end{center}
\end{figure}

{Let us emphasize that 
{the calculations performed by using  the SMC  $F^{p(n)}_2(x)$ \cite{SpinMuonSMC:1997voo} cannot be distinguished by the calculations with our results for $r(x)$ and that also the uncertainties affecting our $r(x)$ for $x>0.85$ (see Fig. \ref{fig:rx}) have no visible effect in Fig. 5.}}

It is rewarding to notice that, for $0.2\le x\le 0.75$, the calculated $^3$He EMC ratio becomes less than 1, without  any 
{dynamical} off-shell correction. This interesting result is generated by
 a clear, although small, combined effect of the careful treatment of the Poincar\'e covariance, the use of nonsymmetric variables to comply with macroscopic locality (cfr. Eq. \eqref{F2e}) and, last but not  least, a realistic description of $^3$He. Moreover, Fig. \ref{fig:EMC} shows
 that the effect of the NNN interaction 
can be safely disregarded for $x\le 0.7$. 
{We also checked that the effects of Coulomb interaction and charge-dependence of the NN Av18 potential are tiny and can be disregarded.}

{Finally, Fig. \ref{fig:sratio} shows the comparison between the super-ratio $R^{ht}(x)$, Eq. \eqref{super}, calculated in our approach and the same quantity extracted by MARATHON  \cite{MARATHON:2021vqu}  by using the theoretical inputs of Refs. \cite{Kulagin:2004ie,Kulagin:2010gd}. 
The black area is generated by our calculations, corresponding to the quadratic  least-squares interpolation of the full dots in Fig. \ref{fig:rx} and the cubic one, respectively.  It   provides  a clear insight  into the {expected} impact of $F^n_2(x)=r(x)~F^p_2(x)$  on the $^3$H SF. It is worth noticing that a super-ratio with a similar pattern, i.e.  definitely bending  for $x>0.7$,  have been extracted i) through a global analysis by  JAM Collaboration \cite{Cocuzza:2021rfn},  ii) by using  off-shell corrections obtained from  Monte-Carlo fits in Ref. \cite{Tropiano:2018quk} (in the same work the super-ratio changes derivative once the off-shell corrections from Ref. \cite{Kulagin:2004ie} are adopted) and iii) in Ref. \cite{Segarra:2019gbp} where    a universal modification of nucleons in short-range correlated  pairs is considered}.

\begin{figure}[htb]
\begin{center}
\includegraphics[width=9. cm]{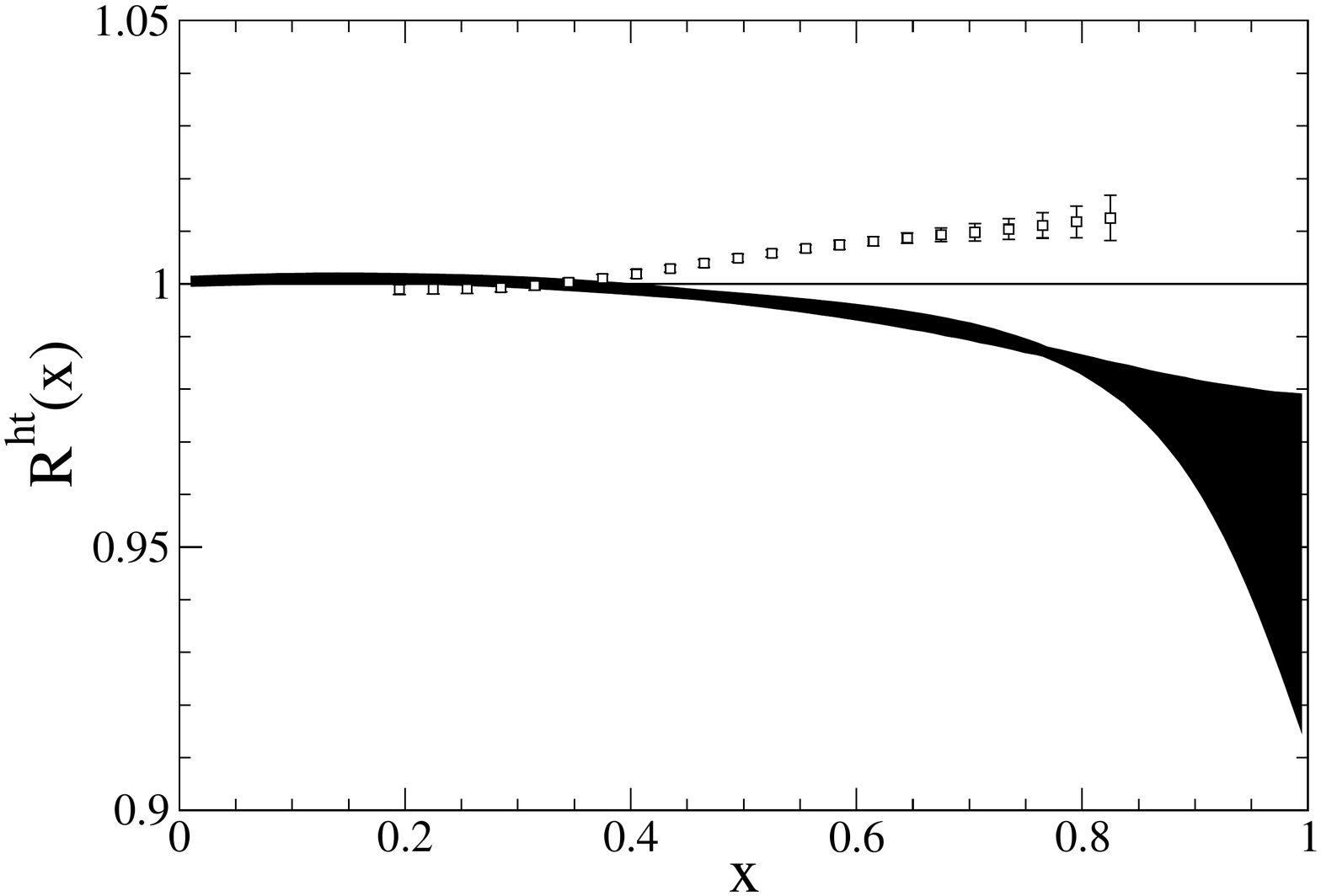}
\caption{The super-ratio $R^{ht}(x)$, Eq. \eqref{super}.  The black area represents our results obtained by using both a quadratic interpolation of the full dots in Fig. \ref{fig:rx} and a cubic one. Empty squares: super-ratio obtained in Ref. \cite{MARATHON:2021vqu} by using the approach of Refs. \cite{Kulagin:2004ie,Kulagin:2010gd} for the theoretical inputs.}
\label{fig:sratio}
\end{center}
\end{figure}

{\it Conclusions.}
A rigorous approach, elaborated within the light-front Hamiltonian dynamics, {satisfying 
{i) Poincar\'e covariance,} ii) the baryon number sum rule, iii) the momentum sum rule and iv) the macroscopic locality,} is applied for  the first time to the description of 
electron inclusive DIS off  {$A=3$ nuclei}, {obtaining the expected signature of the EMC effect}, without medium 
deformations
of the bound nucleons.
{We have adopted} the Bjorken limit, {and used} very accurate 
nuclear wfs~\cite{Kievsky:1994mxj,Kievsky:1995uk}, 
with a rich content of NN and NNN correlations, {also} checking the stability of the result against the use of 
 different 
 nuclear interactions  (NN+NNN or NN only).
 In this way, 
 {within a Poincar\'e-covariant approach,} we are able to set a baseline for starting a consistent separation of the genuinely nuclear effects (where the nucleon dof are acting) from the ones basically pertaining to  QCD. } {with a direct impact on  analysis and interpretation of  future experimental results at the EIC.}

The generalisation to $A>3$ nuclei, where the EMC effect is more pronounced,  e.g. the $^4$He nucleus \cite{Seely:2009gt}, {is straightforward (although numerically challenging)}, since, {in the
Bjorken limit,}
only the 
{momentum space wf} is needed, as shown in Eqs. (\ref{F2a}) and (\ref{F2ab}), valid for any nucleus.  

{\it Acknowledgment.}
M.R. and S.S. thank  for the financial support received under the   STRONG-2020 project of the European Union’s Horizon 2020 research and innovation programme: Fund no 824093.

\newpage
\centerline {\bf Spplemental Material}
 {In Fig. \ref{Eht}, the experimental  ratio~measured by the MARATHON Collaboration \cite{MARATHON:2021vqu}, with very small error bars, and the two  values extrapolated at $x=0$ and $x=1$ of Ref. \cite{Cui:2021gzg}, with large uncertainties, are shown. The solid line is a quadratic interpolation obtained by using a least-squares method. The quadratic parametrization, indicated as  ${\cal {E}}^{ht}(x)$,  is adopted in the recurrence expression of Eq. (12) of  the main text, because of the necessity   to cover the full range of $x$ for evaluating the theoretical super-ratio $ R^{ht}[x,r^{(n)}(x)]$ (cf. Eqs. (11), (10) and (2) in the main text).}
 
 In Fig. 4 of the main text the uncertainties depicted by the shaded area are obtained  by using both a quadratic least-squares interpolation of the full dots, given by
$$r^{q}(x)=1.02107-1.48834~x+0.887864~x^2~,$$   and a cubic one, given by  $$r^{c}(x)= 1.05835-1.99892~x+ 2.31339~x^2
 -1.0557~x^3~.$$ Of course the best interpolation is obtained by using the cubic polynomial.

 \begin{figure}[htb]
\begin{center}
\includegraphics[width=9. cm]{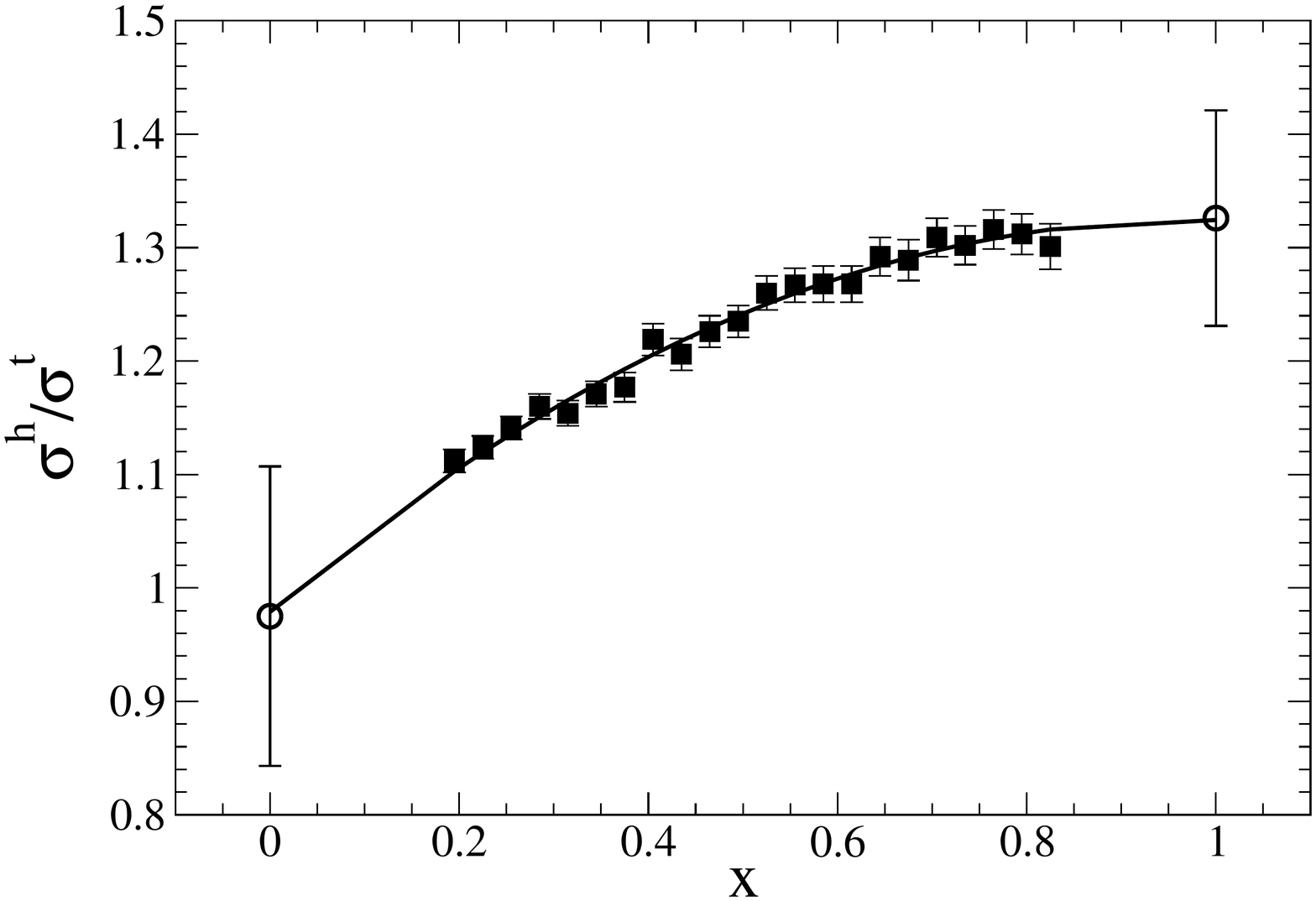}
\caption{The ratio $\sigma^h/ \sigma^t$ of inclusive DIS cross sections off $^3$He and $^3$H.   Black squares: experimental data from Ref. \cite{MARATHON:2021vqu}. Empty circles: extrapolated values   as given in  Ref. \cite{Cui:2021gzg}. Solid line:  quadratic least-squares interpolation.}
\label{Eht}
\end{center}
\end{figure}


\begin{thebibliography}{40}%
\makeatletter
\providecommand \@ifxundefined [1]{%
 \@ifx{#1\undefined}
}%
\providecommand \@ifnum [1]{%
 \ifnum #1\expandafter \@firstoftwo
 \else \expandafter \@secondoftwo
 \fi
}%
\providecommand \@ifx [1]{%
 \ifx #1\expandafter \@firstoftwo
 \else \expandafter \@secondoftwo
 \fi
}%
\providecommand \natexlab [1]{#1}%
\providecommand \enquote  [1]{``#1''}%
\providecommand \bibnamefont  [1]{#1}%
\providecommand \bibfnamefont [1]{#1}%
\providecommand \citenamefont [1]{#1}%
\providecommand \href@noop [0]{\@secondoftwo}%
\providecommand \href [0]{\begingroup \@sanitize@url \@href}%
\providecommand \@href[1]{\@@startlink{#1}\@@href}%
\providecommand \@@href[1]{\endgroup#1\@@endlink}%
\providecommand \@sanitize@url [0]{\catcode `\\12\catcode `\$12\catcode
  `\&12\catcode `\#12\catcode `\^12\catcode `\_12\catcode `\%12\relax}%
\providecommand \@@startlink[1]{}%
\providecommand \@@endlink[0]{}%
\providecommand \url  [0]{\begingroup\@sanitize@url \@url }%
\providecommand \@url [1]{\endgroup\@href {#1}{\urlprefix }}%
\providecommand \urlprefix  [0]{URL }%
\providecommand \Eprint [0]{\href }%
\providecommand \doibase [0]{http://dx.doi.org/}%
\providecommand \selectlanguage [0]{\@gobble}%
\providecommand \bibinfo  [0]{\@secondoftwo}%
\providecommand \bibfield  [0]{\@secondoftwo}%
\providecommand \translation [1]{[#1]}%
\providecommand \BibitemOpen [0]{}%
\providecommand \bibitemStop [0]{}%
\providecommand \bibitemNoStop [0]{.\EOS\space}%
\providecommand \EOS [0]{\spacefactor3000\relax}%
\providecommand \BibitemShut  [1]{\csname bibitem#1\endcsname}%
\let\auto@bib@innerbib\@empty
\bibitem [{\citenamefont {Aubert}\ \emph {et~al.}(1983)\citenamefont {Aubert}
  \emph {et~al.}}]{EuropeanMuon:1983wih}%
  \BibitemOpen
  \bibfield  {author} {\bibinfo {author} {\bibfnamefont {J.~J.}\ \bibnamefont
  {Aubert}} \emph {et~al.} (\bibinfo {collaboration} {European Muon}),\
  }\bibfield  {title} {\enquote {\bibinfo {title} {{The ratio of the nucleon
  structure functions $F_2^n$ for iron and deuterium}},}\ }\href {\doibase
  10.1016/0370-2693(83)90437-9} {\bibfield  {journal} {\bibinfo  {journal}
  {Phys. Lett. B}\ }\textbf {\bibinfo {volume} {123}},\ \bibinfo {pages}
  {275--278} (\bibinfo {year} {1983})}\BibitemShut {NoStop}%
\bibitem [{\citenamefont {Hen}\ \emph {et~al.}(2017)\citenamefont {Hen},
  \citenamefont {Miller}, \citenamefont {Piasetzky},\ and\ \citenamefont
  {Weinstein}}]{Hen:2016kwk}%
  \BibitemOpen
  \bibfield  {author} {\bibinfo {author} {\bibfnamefont {O.}~\bibnamefont
  {Hen}}, \bibinfo {author} {\bibfnamefont {G.~A.}\ \bibnamefont {Miller}},
  \bibinfo {author} {\bibfnamefont {E.}~\bibnamefont {Piasetzky}}, \ and\
  \bibinfo {author} {\bibfnamefont {L.~B.}\ \bibnamefont {Weinstein}},\
  }\bibfield  {title} {\enquote {\bibinfo {title} {{Nucleon-Nucleon
  Correlations, Short-lived Excitations, and the Quarks Within}},}\ }\href
  {\doibase 10.1103/RevModPhys.89.045002} {\bibfield  {journal} {\bibinfo
  {journal} {Rev. Mod. Phys.}\ }\textbf {\bibinfo {volume} {89}},\ \bibinfo
  {pages} {045002} (\bibinfo {year} {2017})},\ \Eprint
  {http://arxiv.org/abs/1611.09748} {arXiv:1611.09748 [nucl-ex]} \BibitemShut
  {NoStop}%
\bibitem [{\citenamefont {Schmookler}\ \emph {et~al.}(2019)\citenamefont
  {Schmookler} \emph {et~al.}}]{CLAS:2019vsb}%
  \BibitemOpen
  \bibfield  {author} {\bibinfo {author} {\bibfnamefont {B.}~\bibnamefont
  {Schmookler}} \emph {et~al.} (\bibinfo {collaboration} {CLAS}),\ }\bibfield
  {title} {\enquote {\bibinfo {title} {{Modified structure of protons and
  neutrons in correlated pairs}},}\ }\href {\doibase 10.1038/s41586-019-0925-9}
  {\bibfield  {journal} {\bibinfo  {journal} {Nature}\ }\textbf {\bibinfo
  {volume} {566}},\ \bibinfo {pages} {354--358} (\bibinfo {year} {2019})},\
  \Eprint {http://arxiv.org/abs/2004.12065} {arXiv:2004.12065 [nucl-ex]}
  \BibitemShut {NoStop}%
\bibitem [{\citenamefont {Wang}\ \emph {et~al.}(2020)\citenamefont {Wang},
  \citenamefont {Thomas},\ and\ \citenamefont {Melnitchouk}}]{Wang:2020uhj}%
  \BibitemOpen
  \bibfield  {author} {\bibinfo {author} {\bibfnamefont {X.~G.}\ \bibnamefont
  {Wang}}, \bibinfo {author} {\bibfnamefont {A.~W.}\ \bibnamefont {Thomas}}, \
  and\ \bibinfo {author} {\bibfnamefont {W.}~\bibnamefont {Melnitchouk}},\
  }\bibfield  {title} {\enquote {\bibinfo {title} {{Do short-range correlations
  cause the nuclear EMC effect in the deuteron?}}}\ }\href {\doibase
  10.1103/PhysRevLett.125.262002} {\bibfield  {journal} {\bibinfo  {journal}
  {Phys. Rev. Lett.}\ }\textbf {\bibinfo {volume} {125}},\ \bibinfo {pages}
  {262002} (\bibinfo {year} {2020})},\ \Eprint
  {http://arxiv.org/abs/2004.03789} {arXiv:2004.03789 [hep-ph]} \BibitemShut
  {NoStop}%
\bibitem [{\citenamefont {Clo\"et}\ \emph {et~al.}(2019)\citenamefont {Clo\"et}
  \emph {et~al.}}]{Cloet:2019mql}%
  \BibitemOpen
  \bibfield  {author} {\bibinfo {author} {\bibfnamefont {I.~C.}\ \bibnamefont
  {Clo\"et}} \emph {et~al.},\ }\bibfield  {title} {\enquote {\bibinfo {title}
  {{Exposing Novel Quark and Gluon Effects in Nuclei}},}\ }\href {\doibase
  10.1088/1361-6471/ab2731} {\bibfield  {journal} {\bibinfo  {journal} {J.
  Phys. G}\ }\textbf {\bibinfo {volume} {46}},\ \bibinfo {pages} {093001}
  (\bibinfo {year} {2019})},\ \Eprint {http://arxiv.org/abs/1902.10572}
  {arXiv:1902.10572 [nucl-ex]} \BibitemShut {NoStop}%
\bibitem [{\citenamefont {Dupr\'e}\ and\ \citenamefont
  {Scopetta}(2016)}]{Dupre:2015jha}%
  \BibitemOpen
  \bibfield  {author} {\bibinfo {author} {\bibfnamefont {R.}~\bibnamefont
  {Dupr\'e}}\ and\ \bibinfo {author} {\bibfnamefont {S.}~\bibnamefont
  {Scopetta}},\ }\bibfield  {title} {\enquote {\bibinfo {title} {{3D Structure
  and Nuclear Targets}},}\ }\href {\doibase 10.1140/epja/i2016-16159-1}
  {\bibfield  {journal} {\bibinfo  {journal} {Eur. Phys. J. A}\ }\textbf
  {\bibinfo {volume} {52}},\ \bibinfo {pages} {159} (\bibinfo {year} {2016})},\
  \Eprint {http://arxiv.org/abs/1510.00794} {arXiv:1510.00794 [nucl-ex]}
  \BibitemShut {NoStop}%
\bibitem [{\citenamefont {Armstrong}\ \emph {et~al.}(2017)\citenamefont
  {Armstrong} \emph {et~al.}}]{Armstrong:2017zqr}%
  \BibitemOpen
  \bibfield  {author} {\bibinfo {author} {\bibfnamefont {W.}~\bibnamefont
  {Armstrong}} \emph {et~al.},\ }\bibfield  {title} {\enquote {\bibinfo {title}
  {{Tagged EMC Measurements on Light Nuclei}},}\ }\href@noop {} {\  (\bibinfo
  {year} {2017})},\ \Eprint {http://arxiv.org/abs/1708.00891} {arXiv:1708.00891
  [nucl-ex]} \BibitemShut {NoStop}%
\bibitem [{\citenamefont {Abdul~Khalek}\ \emph {et~al.}(2021)\citenamefont
  {Abdul~Khalek} \emph {et~al.}}]{AbdulKhalek:2021gbh}%
  \BibitemOpen
  \bibfield  {author} {\bibinfo {author} {\bibfnamefont {R.}~\bibnamefont
  {Abdul~Khalek}} \emph {et~al.},\ }\bibfield  {title} {\enquote {\bibinfo
  {title} {{Science Requirements and Detector Concepts for the Electron-Ion
  Collider: EIC Yellow Report}},}\ }\href@noop {} {\  (\bibinfo {year}
  {2021})},\ \Eprint {http://arxiv.org/abs/2103.05419} {arXiv:2103.05419
  [physics.ins-det]} \BibitemShut {NoStop}%
\bibitem [{\citenamefont {Abrams}\ \emph {et~al.}(2022)\citenamefont {Abrams}
  \emph {et~al.}}]{MARATHON:2021vqu}%
  \BibitemOpen
  \bibfield  {author} {\bibinfo {author} {\bibfnamefont {D.}~\bibnamefont
  {Abrams}} \emph {et~al.} (\bibinfo {collaboration} {MARATHON}),\ }\bibfield
  {title} {\enquote {\bibinfo {title} {{Measurement of the Nucleon
  $F^n_2/F^p_2$ Structure Function Ratio by the Jefferson Lab MARATHON
  Tritium/Helium-3 Deep Inelastic Scattering Experiment}},}\ }\href {\doibase
  10.1103/PhysRevLett.128.132003} {\bibfield  {journal} {\bibinfo  {journal}
  {Phys. Rev. Lett.}\ }\textbf {\bibinfo {volume} {128}},\ \bibinfo {pages}
  {132003} (\bibinfo {year} {2022})},\ \Eprint
  {http://arxiv.org/abs/2104.05850} {arXiv:2104.05850 [hep-ex]} \BibitemShut
  {NoStop}%
\bibitem [{\citenamefont {Cocuzza}\ \emph {et~al.}(2021)\citenamefont
  {Cocuzza}, \citenamefont {Keppel}, \citenamefont {Liu}, \citenamefont
  {Melnitchouk}, \citenamefont {Metz}, \citenamefont {Sato},\ and\
  \citenamefont {Thomas}}]{Cocuzza:2021rfn}%
  \BibitemOpen
  \bibfield  {author} {\bibinfo {author} {\bibfnamefont {C.}~\bibnamefont
  {Cocuzza}}, \bibinfo {author} {\bibfnamefont {C.~E.}\ \bibnamefont {Keppel}},
  \bibinfo {author} {\bibfnamefont {H.}~\bibnamefont {Liu}}, \bibinfo {author}
  {\bibfnamefont {W.}~\bibnamefont {Melnitchouk}}, \bibinfo {author}
  {\bibfnamefont {A.}~\bibnamefont {Metz}}, \bibinfo {author} {\bibfnamefont
  {N.}~\bibnamefont {Sato}}, \ and\ \bibinfo {author} {\bibfnamefont {A.~W.}\
  \bibnamefont {Thomas}} (\bibinfo {collaboration} {Jefferson Lab Angular
  Momentum (JAM)}),\ }\bibfield  {title} {\enquote {\bibinfo {title}
  {{Isovector EMC Effect from Global QCD Analysis with MARATHON Data}},}\
  }\href {\doibase 10.1103/PhysRevLett.127.242001} {\bibfield  {journal}
  {\bibinfo  {journal} {Phys. Rev. Lett.}\ }\textbf {\bibinfo {volume} {127}},\
  \bibinfo {pages} {242001} (\bibinfo {year} {2021})},\ \Eprint
  {http://arxiv.org/abs/2104.06946} {arXiv:2104.06946 [hep-ph]} \BibitemShut
  {NoStop}%
\bibitem [{\citenamefont {Pace}\ \emph {et~al.}(2020)\citenamefont {Pace},
  \citenamefont {Rinaldi}, \citenamefont {Salm\`e},\ and\ \citenamefont
  {Scopetta}}]{Pace:2020ned}%
  \BibitemOpen
  \bibfield  {author} {\bibinfo {author} {\bibfnamefont {Emanuele}\
  \bibnamefont {Pace}}, \bibinfo {author} {\bibfnamefont {Matteo}\ \bibnamefont
  {Rinaldi}}, \bibinfo {author} {\bibfnamefont {Giovanni}\ \bibnamefont
  {Salm\`e}}, \ and\ \bibinfo {author} {\bibfnamefont {Sergio}\ \bibnamefont
  {Scopetta}},\ }\bibfield  {title} {\enquote {\bibinfo {title} {{EMC effect,
  few-nucleon systems and Poincar\'e covariance}},}\ }\href {\doibase
  10.1088/1402-4896/ab8951} {\bibfield  {journal} {\bibinfo  {journal} {Phys.
  Scripta}\ }\textbf {\bibinfo {volume} {95}},\ \bibinfo {pages} {064008}
  (\bibinfo {year} {2020})},\ \Eprint {http://arxiv.org/abs/2004.05877}
  {arXiv:2004.05877 [nucl-th]} \BibitemShut {NoStop}%
\bibitem [{\citenamefont {Dirac}(1949)}]{Dirac:1949cp}%
  \BibitemOpen
  \bibfield  {author} {\bibinfo {author} {\bibfnamefont {Paul A.~M.}\
  \bibnamefont {Dirac}},\ }\bibfield  {title} {\enquote {\bibinfo {title}
  {{Forms of Relativistic Dynamics}},}\ }\href {\doibase
  10.1103/RevModPhys.21.392} {\bibfield  {journal} {\bibinfo  {journal} {Rev.
  Mod. Phys.}\ }\textbf {\bibinfo {volume} {21}},\ \bibinfo {pages} {392--399}
  (\bibinfo {year} {1949})}\BibitemShut {NoStop}%
\bibitem [{\citenamefont {Keister}\ and\ \citenamefont {Polyzou}(1991)}]{KP}%
  \BibitemOpen
  \bibfield  {author} {\bibinfo {author} {\bibfnamefont {B.~D.}\ \bibnamefont
  {Keister}}\ and\ \bibinfo {author} {\bibfnamefont {W.~N.}\ \bibnamefont
  {Polyzou}},\ }\bibfield  {title} {\enquote {\bibinfo {title} {{Relativistic
  Hamiltonian dynamics in nuclear and particle physics}},}\ }\href@noop {}
  {\bibfield  {journal} {\bibinfo  {journal} {Adv. Nucl. Phys.}\ }\textbf
  {\bibinfo {volume} {20}},\ \bibinfo {pages} {225--479} (\bibinfo {year}
  {1991})}\BibitemShut {NoStop}%
\bibitem [{\citenamefont {Del~Dotto}\ \emph {et~al.}(2017)\citenamefont
  {Del~Dotto}, \citenamefont {Pace}, \citenamefont {Salm\`e},\ and\
  \citenamefont {Scopetta}}]{DelDotto:2016vkh}%
  \BibitemOpen
  \bibfield  {author} {\bibinfo {author} {\bibfnamefont {Alessio}\ \bibnamefont
  {Del~Dotto}}, \bibinfo {author} {\bibfnamefont {Emanuele}\ \bibnamefont
  {Pace}}, \bibinfo {author} {\bibfnamefont {Giovanni}\ \bibnamefont
  {Salm\`e}}, \ and\ \bibinfo {author} {\bibfnamefont {Sergio}\ \bibnamefont
  {Scopetta}},\ }\bibfield  {title} {\enquote {\bibinfo {title} {{Light-Front
  spin-dependent Spectral Function and Nucleon Momentum Distributions for a
  Three-Body System}},}\ }\href {\doibase 10.1103/PhysRevC.95.014001}
  {\bibfield  {journal} {\bibinfo  {journal} {Phys. Rev. C}\ }\textbf {\bibinfo
  {volume} {95}},\ \bibinfo {pages} {014001} (\bibinfo {year} {2017})},\
  \Eprint {http://arxiv.org/abs/1609.03804} {arXiv:1609.03804 [nucl-th]}
  \BibitemShut {NoStop}%
\bibitem [{\citenamefont {Alessandro}\ \emph {et~al.}(2021)\citenamefont
  {Alessandro}, \citenamefont {Del~Dotto}, \citenamefont {Pace}, \citenamefont
  {Perna}, \citenamefont {Salm\`e},\ and\ \citenamefont
  {Scopetta}}]{Alessandro:2021cbg}%
  \BibitemOpen
  \bibfield  {author} {\bibinfo {author} {\bibfnamefont {Rocco}\ \bibnamefont
  {Alessandro}}, \bibinfo {author} {\bibfnamefont {Alessio}\ \bibnamefont
  {Del~Dotto}}, \bibinfo {author} {\bibfnamefont {Emanuele}\ \bibnamefont
  {Pace}}, \bibinfo {author} {\bibfnamefont {Gabriele}\ \bibnamefont {Perna}},
  \bibinfo {author} {\bibfnamefont {Giovanni}\ \bibnamefont {Salm\`e}}, \ and\
  \bibinfo {author} {\bibfnamefont {Sergio}\ \bibnamefont {Scopetta}},\
  }\bibfield  {title} {\enquote {\bibinfo {title} {{Light-front transverse
  momentum distributions for J=1/2 hadronic systems in valence
  approximation}},}\ }\href {\doibase 10.1103/PhysRevC.104.065204} {\bibfield
  {journal} {\bibinfo  {journal} {Phys. Rev. C}\ }\textbf {\bibinfo {volume}
  {104}},\ \bibinfo {pages} {065204} (\bibinfo {year} {2021})},\ \Eprint
  {http://arxiv.org/abs/2107.10187} {arXiv:2107.10187 [nucl-th]} \BibitemShut
  {NoStop}%
\bibitem [{\citenamefont {Pace}\ \emph
  {et~al.}(2001{\natexlab{a}})\citenamefont {Pace}, \citenamefont {Salm\`e},\
  and\ \citenamefont {Scopetta}}]{Pace:2000ky}%
  \BibitemOpen
  \bibfield  {author} {\bibinfo {author} {\bibfnamefont {E.}~\bibnamefont
  {Pace}}, \bibinfo {author} {\bibfnamefont {G.}~\bibnamefont {Salm\`e}}, \
  and\ \bibinfo {author} {\bibfnamefont {S.}~\bibnamefont {Scopetta}},\
  }\bibfield  {title} {\enquote {\bibinfo {title} {{Neutron unpolarized
  structure function $F^N_2(x)$ from deep inelastic scattering off $^3$He and
  $^3$H}},}\ }\href {\doibase 10.1016/S0375-9474(01)00877-6} {\bibfield
  {journal} {\bibinfo  {journal} {Nucl. Phys. A}\ }\textbf {\bibinfo {volume}
  {689}},\ \bibinfo {pages} {453--456} (\bibinfo {year}
  {2001}{\natexlab{a}})},\ \Eprint {http://arxiv.org/abs/nucl-th/0009028}
  {arXiv:nucl-th/0009028} \BibitemShut {NoStop}%
\bibitem [{\citenamefont {Pace}\ \emph
  {et~al.}(2001{\natexlab{b}})\citenamefont {Pace}, \citenamefont {Salm\`e},
  \citenamefont {Scopetta},\ and\ \citenamefont {Kievsky}}]{Pace:2001cm}%
  \BibitemOpen
  \bibfield  {author} {\bibinfo {author} {\bibfnamefont {E.}~\bibnamefont
  {Pace}}, \bibinfo {author} {\bibfnamefont {G.}~\bibnamefont {Salm\`e}},
  \bibinfo {author} {\bibfnamefont {S.}~\bibnamefont {Scopetta}}, \ and\
  \bibinfo {author} {\bibfnamefont {A.}~\bibnamefont {Kievsky}},\ }\bibfield
  {title} {\enquote {\bibinfo {title} {{Neutron structure function F(2)**n (x)
  from deep inelastic electron scattering off few nucleon systems}},}\ }\href
  {\doibase 10.1103/PhysRevC.64.055203} {\bibfield  {journal} {\bibinfo
  {journal} {Phys. Rev. C}\ }\textbf {\bibinfo {volume} {64}},\ \bibinfo
  {pages} {055203} (\bibinfo {year} {2001}{\natexlab{b}})},\ \Eprint
  {http://arxiv.org/abs/nucl-th/0109005} {arXiv:nucl-th/0109005} \BibitemShut
  {NoStop}%
\bibitem [{\citenamefont {Adeva}\ \emph {et~al.}(1997)\citenamefont {Adeva}
  \emph {et~al.}}]{SpinMuonSMC:1997voo}%
  \BibitemOpen
  \bibfield  {author} {\bibinfo {author} {\bibfnamefont {B.}~\bibnamefont
  {Adeva}} \emph {et~al.} (\bibinfo {collaboration} {Spin Muon Collab.
  (SMC)}),\ }\bibfield  {title} {\enquote {\bibinfo {title} {{The Spin
  dependent structure function $g_1 (x)$ of the proton from polarized deep
  inelastic muon scattering}},}\ }\href {\doibase
  10.1016/S0370-2693(97)01106-4} {\bibfield  {journal} {\bibinfo  {journal}
  {Phys. Lett. B}\ }\textbf {\bibinfo {volume} {412}},\ \bibinfo {pages}
  {414--424} (\bibinfo {year} {1997})}\BibitemShut {NoStop}%
\bibitem [{\citenamefont {Smith}\ and\ \citenamefont
  {Miller}(2002)}]{Smith:2002ci}%
  \BibitemOpen
  \bibfield  {author} {\bibinfo {author} {\bibfnamefont {Jason~Robert}\
  \bibnamefont {Smith}}\ and\ \bibinfo {author} {\bibfnamefont {Gerald~A.}\
  \bibnamefont {Miller}},\ }\bibfield  {title} {\enquote {\bibinfo {title}
  {{Return of the EMC effect: Finite nuclei}},}\ }\href {\doibase
  10.1103/PhysRevC.65.055206} {\bibfield  {journal} {\bibinfo  {journal} {Phys.
  Rev. C}\ }\textbf {\bibinfo {volume} {65}},\ \bibinfo {pages} {055206}
  (\bibinfo {year} {2002})},\ \Eprint {http://arxiv.org/abs/nucl-th/0202016}
  {arXiv:nucl-th/0202016} \BibitemShut {NoStop}%
\bibitem [{\citenamefont {Miller}\ and\ \citenamefont
  {Smith}(2002)}]{Miller:2001tg}%
  \BibitemOpen
  \bibfield  {author} {\bibinfo {author} {\bibfnamefont {Gerald~A.}\
  \bibnamefont {Miller}}\ and\ \bibinfo {author} {\bibfnamefont {Jason~Robert}\
  \bibnamefont {Smith}},\ }\bibfield  {title} {\enquote {\bibinfo {title}
  {{Return of the EMC effect}},}\ }\href {\doibase 10.1103/PhysRevC.66.049903}
  {\bibfield  {journal} {\bibinfo  {journal} {Phys. Rev. C}\ }\textbf {\bibinfo
  {volume} {65}},\ \bibinfo {pages} {015211} (\bibinfo {year} {2002})},\
  \bibinfo {note} {[Erratum: Phys.Rev.C 66, 049903 (2002)]},\ \Eprint
  {http://arxiv.org/abs/nucl-th/0107026} {arXiv:nucl-th/0107026} \BibitemShut
  {NoStop}%
\bibitem [{\citenamefont {Oelfke}\ \emph {et~al.}(1990)\citenamefont {Oelfke},
  \citenamefont {Sauer},\ and\ \citenamefont {Coester}}]{Oelfke:1990uy}%
  \BibitemOpen
  \bibfield  {author} {\bibinfo {author} {\bibfnamefont {U.}~\bibnamefont
  {Oelfke}}, \bibinfo {author} {\bibfnamefont {P.~U.}\ \bibnamefont {Sauer}}, \
  and\ \bibinfo {author} {\bibfnamefont {F.}~\bibnamefont {Coester}},\
  }\bibfield  {title} {\enquote {\bibinfo {title} {{Convolution models of deep
  inelastic scattering: The Three nucleon bound states as a test case}},}\
  }\href {\doibase 10.1016/0375-9474(90)90181-K} {\bibfield  {journal}
  {\bibinfo  {journal} {Nucl. Phys. A}\ }\textbf {\bibinfo {volume} {518}},\
  \bibinfo {pages} {593--616} (\bibinfo {year} {1990})}\BibitemShut {NoStop}%
\bibitem [{\citenamefont {Coester}(1992)}]{Coester:1992cg}%
  \BibitemOpen
  \bibfield  {author} {\bibinfo {author} {\bibfnamefont {F.}~\bibnamefont
  {Coester}},\ }\bibfield  {title} {\enquote {\bibinfo {title} {{Null plane
  dynamics of particles and fields}},}\ }\href {\doibase
  10.1016/0146-6410(92)90002-J} {\bibfield  {journal} {\bibinfo  {journal}
  {Prog. Part. Nucl. Phys.}\ }\textbf {\bibinfo {volume} {29}},\ \bibinfo
  {pages} {1--32} (\bibinfo {year} {1992})}\BibitemShut {NoStop}%
\bibitem [{\citenamefont {Kievsky}\ \emph {et~al.}(1994)\citenamefont
  {Kievsky}, \citenamefont {Viviani},\ and\ \citenamefont
  {Rosati}}]{Kievsky:1994mxj}%
  \BibitemOpen
  \bibfield  {author} {\bibinfo {author} {\bibfnamefont {A.}~\bibnamefont
  {Kievsky}}, \bibinfo {author} {\bibfnamefont {M.}~\bibnamefont {Viviani}}, \
  and\ \bibinfo {author} {\bibfnamefont {S.}~\bibnamefont {Rosati}},\
  }\bibfield  {title} {\enquote {\bibinfo {title} {{Study of bound and
  scattering states of three nucleon systems}},}\ }\href {\doibase
  10.1016/0375-9474(94)90931-8} {\bibfield  {journal} {\bibinfo  {journal}
  {Nucl. Phys. A}\ }\textbf {\bibinfo {volume} {577}},\ \bibinfo {pages}
  {511--527} (\bibinfo {year} {1994})},\ \Eprint
  {http://arxiv.org/abs/nucl-th/9706067} {arXiv:nucl-th/9706067} \BibitemShut
  {NoStop}%
\bibitem [{\citenamefont {Kievsky}\ \emph {et~al.}(1995)\citenamefont
  {Kievsky}, \citenamefont {Viviani},\ and\ \citenamefont
  {Rosati}}]{Kievsky:1995uk}%
  \BibitemOpen
  \bibfield  {author} {\bibinfo {author} {\bibfnamefont {A.}~\bibnamefont
  {Kievsky}}, \bibinfo {author} {\bibfnamefont {M.}~\bibnamefont {Viviani}}, \
  and\ \bibinfo {author} {\bibfnamefont {S.}~\bibnamefont {Rosati}},\
  }\bibfield  {title} {\enquote {\bibinfo {title} {{Cross-section, polarization
  observables, and phase shift parameters in proton deuteron and neutron
  deuteron elastic scattering}},}\ }\href {\doibase 10.1103/PhysRevC.52.R15}
  {\bibfield  {journal} {\bibinfo  {journal} {Phys. Rev. C}\ }\textbf {\bibinfo
  {volume} {52}},\ \bibinfo {pages} {R15--R19} (\bibinfo {year}
  {1995})}\BibitemShut {NoStop}%
\bibitem [{\citenamefont {Wiringa}\ \emph {et~al.}(1995)\citenamefont
  {Wiringa}, \citenamefont {Stoks},\ and\ \citenamefont
  {Schiavilla}}]{Wiringa:1994wb}%
  \BibitemOpen
  \bibfield  {author} {\bibinfo {author} {\bibfnamefont {Robert~B.}\
  \bibnamefont {Wiringa}}, \bibinfo {author} {\bibfnamefont {V.~G.~J.}\
  \bibnamefont {Stoks}}, \ and\ \bibinfo {author} {\bibfnamefont
  {R.}~\bibnamefont {Schiavilla}},\ }\bibfield  {title} {\enquote {\bibinfo
  {title} {{An Accurate nucleon-nucleon potential with charge independence
  breaking}},}\ }\href {\doibase 10.1103/PhysRevC.51.38} {\bibfield  {journal}
  {\bibinfo  {journal} {Phys. Rev. C}\ }\textbf {\bibinfo {volume} {51}},\
  \bibinfo {pages} {38--51} (\bibinfo {year} {1995})},\ \Eprint
  {http://arxiv.org/abs/nucl-th/9408016} {arXiv:nucl-th/9408016} \BibitemShut
  {NoStop}%
\bibitem [{\citenamefont {Pudliner}\ \emph {et~al.}(1995)\citenamefont
  {Pudliner}, \citenamefont {Pandharipande}, \citenamefont {Carlson},\ and\
  \citenamefont {Wiringa}}]{Pudliner:1995wk}%
  \BibitemOpen
  \bibfield  {author} {\bibinfo {author} {\bibfnamefont {B.~S.}\ \bibnamefont
  {Pudliner}}, \bibinfo {author} {\bibfnamefont {V.~R.}\ \bibnamefont
  {Pandharipande}}, \bibinfo {author} {\bibfnamefont {J.}~\bibnamefont
  {Carlson}}, \ and\ \bibinfo {author} {\bibfnamefont {Robert~B.}\ \bibnamefont
  {Wiringa}},\ }\bibfield  {title} {\enquote {\bibinfo {title} {{Quantum Monte
  Carlo calculations of A $\le$ 6 nuclei}},}\ }\href {\doibase
  10.1103/PhysRevLett.74.4396} {\bibfield  {journal} {\bibinfo  {journal}
  {Phys. Rev. Lett.}\ }\textbf {\bibinfo {volume} {74}},\ \bibinfo {pages}
  {4396--4399} (\bibinfo {year} {1995})},\ \Eprint
  {http://arxiv.org/abs/nucl-th/9502031} {arXiv:nucl-th/9502031} \BibitemShut
  {NoStop}%
\bibitem [{\citenamefont {Lev}\ \emph {et~al.}(1998)\citenamefont {Lev},
  \citenamefont {Pace},\ and\ \citenamefont {Salm\`e}}]{Lev:1998qz}%
  \BibitemOpen
  \bibfield  {author} {\bibinfo {author} {\bibfnamefont {F.~M.}\ \bibnamefont
  {Lev}}, \bibinfo {author} {\bibfnamefont {E.}~\bibnamefont {Pace}}, \ and\
  \bibinfo {author} {\bibfnamefont {G.}~\bibnamefont {Salm\`e}},\ }\bibfield
  {title} {\enquote {\bibinfo {title} {{Electromagnetic and weak current
  operators for interacting systems within the front form dynamics}},}\ }\href
  {\doibase 10.1016/S0375-9474(98)00469-2} {\bibfield  {journal} {\bibinfo
  {journal} {Nucl. Phys. A}\ }\textbf {\bibinfo {volume} {641}},\ \bibinfo
  {pages} {229--259} (\bibinfo {year} {1998})},\ \Eprint
  {http://arxiv.org/abs/hep-ph/9807255} {arXiv:hep-ph/9807255} \BibitemShut
  {NoStop}%
\bibitem [{\citenamefont {Ciofi~degli Atti}\ \emph {et~al.}(1993)\citenamefont
  {Ciofi~degli Atti}, \citenamefont {Scopetta}, \citenamefont {Pace},\ and\
  \citenamefont {Salm\`e}}]{CiofidegliAtti:1993zs}%
  \BibitemOpen
  \bibfield  {author} {\bibinfo {author} {\bibfnamefont {Claudio}\ \bibnamefont
  {Ciofi~degli Atti}}, \bibinfo {author} {\bibfnamefont {S.}~\bibnamefont
  {Scopetta}}, \bibinfo {author} {\bibfnamefont {E.}~\bibnamefont {Pace}}, \
  and\ \bibinfo {author} {\bibfnamefont {G.}~\bibnamefont {Salm\`e}},\
  }\bibfield  {title} {\enquote {\bibinfo {title} {{Nuclear effects in deep
  inelastic scattering of polarized electrons off polarized He-3 and the
  neutron spin structure functions}},}\ }\href {\doibase
  10.1103/PhysRevC.48.R968} {\bibfield  {journal} {\bibinfo  {journal} {Phys.
  Rev. C}\ }\textbf {\bibinfo {volume} {48}},\ \bibinfo {pages} {R968--R972}
  (\bibinfo {year} {1993})},\ \Eprint {http://arxiv.org/abs/nucl-th/9303016}
  {arXiv:nucl-th/9303016} \BibitemShut {NoStop}%
\bibitem [{\citenamefont {Scopetta}(2007)}]{Scopetta:2006ww}%
  \BibitemOpen
  \bibfield  {author} {\bibinfo {author} {\bibfnamefont {Sergio}\ \bibnamefont
  {Scopetta}},\ }\bibfield  {title} {\enquote {\bibinfo {title} {{Neutron
  single spin asymmetries from semi-inclusive deep inelastic scattering off
  transversely polarized He-3}},}\ }\href {\doibase 10.1103/PhysRevD.75.054005}
  {\bibfield  {journal} {\bibinfo  {journal} {Phys. Rev. D}\ }\textbf {\bibinfo
  {volume} {75}},\ \bibinfo {pages} {054005} (\bibinfo {year} {2007})},\
  \Eprint {http://arxiv.org/abs/hep-ph/0612354} {arXiv:hep-ph/0612354}
  \BibitemShut {NoStop}%
\bibitem [{\citenamefont {Kaptari}\ \emph {et~al.}(2014)\citenamefont
  {Kaptari}, \citenamefont {Del~Dotto}, \citenamefont {Pace}, \citenamefont
  {Salm\`e},\ and\ \citenamefont {Scopetta}}]{Kaptari:2013dma}%
  \BibitemOpen
  \bibfield  {author} {\bibinfo {author} {\bibfnamefont {L.~P.}\ \bibnamefont
  {Kaptari}}, \bibinfo {author} {\bibfnamefont {A.}~\bibnamefont {Del~Dotto}},
  \bibinfo {author} {\bibfnamefont {E.}~\bibnamefont {Pace}}, \bibinfo {author}
  {\bibfnamefont {G.}~\bibnamefont {Salm\`e}}, \ and\ \bibinfo {author}
  {\bibfnamefont {S.}~\bibnamefont {Scopetta}},\ }\bibfield  {title} {\enquote
  {\bibinfo {title} {{Distorted spin-dependent spectral function of an A=3
  nucleus and semi-inclusive deep inelastic scattering processes}},}\ }\href
  {\doibase 10.1103/PhysRevC.89.035206} {\bibfield  {journal} {\bibinfo
  {journal} {Phys. Rev. C}\ }\textbf {\bibinfo {volume} {89}},\ \bibinfo
  {pages} {035206} (\bibinfo {year} {2014})},\ \Eprint
  {http://arxiv.org/abs/1307.2848} {arXiv:1307.2848 [nucl-th]} \BibitemShut
  {NoStop}%
\bibitem [{\citenamefont {Aubert}\ \emph {et~al.}(1987)\citenamefont {Aubert}
  \emph {et~al.}}]{EuropeanMuon:1987obv}%
  \BibitemOpen
  \bibfield  {author} {\bibinfo {author} {\bibfnamefont {J.~J.}\ \bibnamefont
  {Aubert}} \emph {et~al.} (\bibinfo {collaboration} {European Muon}),\
  }\bibfield  {title} {\enquote {\bibinfo {title} {{Measurements of the nucleon
  structure functions $F2_n$ in deep inelastic muon scattering from deuterium
  and comparison with those from hydrogen and iron}},}\ }\href {\doibase
  10.1016/0550-3213(87)90090-3} {\bibfield  {journal} {\bibinfo  {journal}
  {Nucl. Phys. B}\ }\textbf {\bibinfo {volume} {293}},\ \bibinfo {pages}
  {740--786} (\bibinfo {year} {1987})}\BibitemShut {NoStop}%
\bibitem [{\citenamefont {Gluck}\ \emph {et~al.}(1992)\citenamefont {Gluck},
  \citenamefont {Reya},\ and\ \citenamefont {Vogt}}]{Gluck:1991ng}%
  \BibitemOpen
  \bibfield  {author} {\bibinfo {author} {\bibfnamefont {M.}~\bibnamefont
  {Gluck}}, \bibinfo {author} {\bibfnamefont {E.}~\bibnamefont {Reya}}, \ and\
  \bibinfo {author} {\bibfnamefont {A.}~\bibnamefont {Vogt}},\ }\bibfield
  {title} {\enquote {\bibinfo {title} {{Parton distributions for high-energy
  collisions}},}\ }\href {\doibase 10.1007/BF01483880} {\bibfield  {journal}
  {\bibinfo  {journal} {Z. Phys. C}\ }\textbf {\bibinfo {volume} {53}},\
  \bibinfo {pages} {127--134} (\bibinfo {year} {1992})}\BibitemShut {NoStop}%
\bibitem [{\citenamefont {Martin}\ \emph {et~al.}(2002)\citenamefont {Martin},
  \citenamefont {Roberts}, \citenamefont {Stirling},\ and\ \citenamefont
  {Thorne}}]{Martin:2001es}%
  \BibitemOpen
  \bibfield  {author} {\bibinfo {author} {\bibfnamefont {Alan~D.}\ \bibnamefont
  {Martin}}, \bibinfo {author} {\bibfnamefont {R.~G.}\ \bibnamefont {Roberts}},
  \bibinfo {author} {\bibfnamefont {W.~J.}\ \bibnamefont {Stirling}}, \ and\
  \bibinfo {author} {\bibfnamefont {R.~S.}\ \bibnamefont {Thorne}},\ }\bibfield
   {title} {\enquote {\bibinfo {title} {{MRST2001: Partons and $\alpha_s$ from
  precise deep inelastic scattering and Tevatron jet data}},}\ }\href {\doibase
  10.1007/s100520100842} {\bibfield  {journal} {\bibinfo  {journal} {Eur. Phys.
  J. C}\ }\textbf {\bibinfo {volume} {23}},\ \bibinfo {pages} {73--87}
  (\bibinfo {year} {2002})},\ \Eprint {http://arxiv.org/abs/hep-ph/0110215}
  {arXiv:hep-ph/0110215} \BibitemShut {NoStop}%
\bibitem [{\citenamefont {Cui}\ \emph {et~al.}(2022)\citenamefont {Cui},
  \citenamefont {Gao}, \citenamefont {Binosi}, \citenamefont {Chang},
  \citenamefont {Roberts},\ and\ \citenamefont {Schmidt}}]{Cui:2021gzg}%
  \BibitemOpen
  \bibfield  {author} {\bibinfo {author} {\bibfnamefont {Zhu-Fang}\
  \bibnamefont {Cui}}, \bibinfo {author} {\bibfnamefont {Fei}\ \bibnamefont
  {Gao}}, \bibinfo {author} {\bibfnamefont {Daniele}\ \bibnamefont {Binosi}},
  \bibinfo {author} {\bibfnamefont {Lei}\ \bibnamefont {Chang}}, \bibinfo
  {author} {\bibfnamefont {Craig~D.}\ \bibnamefont {Roberts}}, \ and\ \bibinfo
  {author} {\bibfnamefont {Sebastian~M.}\ \bibnamefont {Schmidt}},\ }\bibfield
  {title} {\enquote {\bibinfo {title} {{Valence Quark Ratio in the Proton}},}\
  }\href {\doibase 10.1088/0256-307X/39/4/041401} {\bibfield  {journal}
  {\bibinfo  {journal} {Chin. Phys. Lett.}\ }\textbf {\bibinfo {volume} {39}},\
  \bibinfo {pages} {041401} (\bibinfo {year} {2022})},\ \Eprint
  {http://arxiv.org/abs/2108.11493} {arXiv:2108.11493 [hep-ph]} \BibitemShut
  {NoStop}%
\bibitem [{\citenamefont {Kulagin}\ and\ \citenamefont
  {Petti}(2006)}]{Kulagin:2004ie}%
  \BibitemOpen
  \bibfield  {author} {\bibinfo {author} {\bibfnamefont {Sergey~A.}\
  \bibnamefont {Kulagin}}\ and\ \bibinfo {author} {\bibfnamefont
  {R.}~\bibnamefont {Petti}},\ }\bibfield  {title} {\enquote {\bibinfo {title}
  {{Global study of nuclear structure functions}},}\ }\href {\doibase
  10.1016/j.nuclphysa.2005.10.011} {\bibfield  {journal} {\bibinfo  {journal}
  {Nucl. Phys. A}\ }\textbf {\bibinfo {volume} {765}},\ \bibinfo {pages}
  {126--187} (\bibinfo {year} {2006})},\ \Eprint
  {http://arxiv.org/abs/hep-ph/0412425} {arXiv:hep-ph/0412425} \BibitemShut
  {NoStop}%
\bibitem [{\citenamefont {Kulagin}\ and\ \citenamefont
  {Petti}(2010)}]{Kulagin:2010gd}%
  \BibitemOpen
  \bibfield  {author} {\bibinfo {author} {\bibfnamefont {S.~A.}\ \bibnamefont
  {Kulagin}}\ and\ \bibinfo {author} {\bibfnamefont {R.}~\bibnamefont
  {Petti}},\ }\bibfield  {title} {\enquote {\bibinfo {title} {{Structure
  functions for light nuclei}},}\ }\href {\doibase 10.1103/PhysRevC.82.054614}
  {\bibfield  {journal} {\bibinfo  {journal} {Phys. Rev. C}\ }\textbf {\bibinfo
  {volume} {82}},\ \bibinfo {pages} {054614} (\bibinfo {year} {2010})},\
  \Eprint {http://arxiv.org/abs/1004.3062} {arXiv:1004.3062 [hep-ph]}
  \BibitemShut {NoStop}%
\bibitem [{\citenamefont {Seely}\ \emph {et~al.}(2009)\citenamefont {Seely}
  \emph {et~al.}}]{Seely:2009gt}%
  \BibitemOpen
  \bibfield  {author} {\bibinfo {author} {\bibfnamefont {J.}~\bibnamefont
  {Seely}} \emph {et~al.},\ }\bibfield  {title} {\enquote {\bibinfo {title}
  {{New measurements of the EMC effect in very light nuclei}},}\ }\href
  {\doibase 10.1103/PhysRevLett.103.202301} {\bibfield  {journal} {\bibinfo
  {journal} {Phys. Rev. Lett.}\ }\textbf {\bibinfo {volume} {103}},\ \bibinfo
  {pages} {202301} (\bibinfo {year} {2009})},\ \Eprint
  {http://arxiv.org/abs/0904.4448} {arXiv:0904.4448 [nucl-ex]} \BibitemShut
  {NoStop}%
\bibitem [{\citenamefont {Pace}\ and\ \citenamefont
  {Salm\`e}(2001)}]{Pace:2001vg}%
  \BibitemOpen
  \bibfield  {author} {\bibinfo {author} {\bibfnamefont {E.}~\bibnamefont
  {Pace}}\ and\ \bibinfo {author} {\bibfnamefont {G.}~\bibnamefont {Salm\`e}},\
  }\bibfield  {title} {\enquote {\bibinfo {title} {{Deuteron electromagnetic
  properties with a Poincare covariant current operator within front form
  Hamiltonian dynamics}},}\ }in\ \href {\doibase 10.1142/9789812811356_0028}
  {\emph {\bibinfo {booktitle} {{8th Conference on Problems in Theoretical
  Nuclear Physics}}}}\ (\bibinfo {year} {2001})\ \Eprint
  {http://arxiv.org/abs/nucl-th/0106004} {arXiv:nucl-th/0106004} \BibitemShut
  {NoStop}%
\bibitem [{\citenamefont {Tropiano}\ \emph {et~al.}(2019)\citenamefont
  {Tropiano}, \citenamefont {Ethier}, \citenamefont {Melnitchouk},\ and\
  \citenamefont {Sato}}]{Tropiano:2018quk}%
  \BibitemOpen
  \bibfield  {author} {\bibinfo {author} {\bibfnamefont {A.~J.}\ \bibnamefont
  {Tropiano}}, \bibinfo {author} {\bibfnamefont {J.~J.}\ \bibnamefont
  {Ethier}}, \bibinfo {author} {\bibfnamefont {W.}~\bibnamefont {Melnitchouk}},
  \ and\ \bibinfo {author} {\bibfnamefont {N.}~\bibnamefont {Sato}},\
  }\bibfield  {title} {\enquote {\bibinfo {title} {{Deep-inelastic and
  quasielastic electron scattering from $A=3$ nuclei}},}\ }\href {\doibase
  10.1103/PhysRevC.99.035201} {\bibfield  {journal} {\bibinfo  {journal} {Phys.
  Rev. C}\ }\textbf {\bibinfo {volume} {99}},\ \bibinfo {pages} {035201}
  (\bibinfo {year} {2019})},\ \Eprint {http://arxiv.org/abs/1811.07668}
  {arXiv:1811.07668 [nucl-th]} \BibitemShut {NoStop}%
\bibitem [{\citenamefont {Segarra}\ \emph {et~al.}(2020)\citenamefont
  {Segarra}, \citenamefont {Schmidt}, \citenamefont {Kutz}, \citenamefont
  {Higinbotham}, \citenamefont {Piasetzky}, \citenamefont {Strikman},
  \citenamefont {Weinstein},\ and\ \citenamefont {Hen}}]{Segarra:2019gbp}%
  \BibitemOpen
  \bibfield  {author} {\bibinfo {author} {\bibfnamefont {E.~P.}\ \bibnamefont
  {Segarra}}, \bibinfo {author} {\bibfnamefont {A.}~\bibnamefont {Schmidt}},
  \bibinfo {author} {\bibfnamefont {T.}~\bibnamefont {Kutz}}, \bibinfo {author}
  {\bibfnamefont {D.~W.}\ \bibnamefont {Higinbotham}}, \bibinfo {author}
  {\bibfnamefont {E.}~\bibnamefont {Piasetzky}}, \bibinfo {author}
  {\bibfnamefont {M.}~\bibnamefont {Strikman}}, \bibinfo {author}
  {\bibfnamefont {L.~B.}\ \bibnamefont {Weinstein}}, \ and\ \bibinfo {author}
  {\bibfnamefont {O.}~\bibnamefont {Hen}},\ }\bibfield  {title} {\enquote
  {\bibinfo {title} {{Neutron Valence Structure from Nuclear Deep Inelastic
  Scattering}},}\ }\href {\doibase 10.1103/PhysRevLett.124.092002} {\bibfield
  {journal} {\bibinfo  {journal} {Phys. Rev. Lett.}\ }\textbf {\bibinfo
  {volume} {124}},\ \bibinfo {pages} {092002} (\bibinfo {year} {2020})},\
  \Eprint {http://arxiv.org/abs/1908.02223} {arXiv:1908.02223 [nucl-th]}
  \BibitemShut {NoStop}%
\end{thebibliography}

%

\end{document}